\documentclass{article}              

\usepackage{rikrutjens}
\usepackage{authblk}
\usepackage{subfigure}
\usepackage{subfloat}

\begin{document}

\title{Method of Distributions for 
Systems with Stochastic Forcing}
\author[1,3]{Rik J.L. Rutjens}
\author[1]{Gustaaf B. Jacobs\thanks{gjacobs@sdsu.edu}}
\author[2]{Daniel M. Tartakovsky}
\affil[1]{Department of Aerospace Engineering, San Diego State University, San Diego, CA 92182, USA}
\affil[2]{Department of Energy Resources Engineering, Stanford University, Stanford, CA 94305, USA}
\affil[3]{Department of Mathematics and Computer Science, Eindhoven University of Technology, Eindhoven, The Netherlands}


\maketitle

\begin{abstract}
The method of distributions is developed for systems that are governed by
	hyperbolic conservation laws  with stochastic forcing. The method
	yields a deterministic equation for the cumulative density distribution
	(CDF) of a system state, e.g., for flow velocity governed by an
	inviscid  Burgers' equation with random source coefficients. This is
	achieved without recourse to any closure approximation. The CDF model
	is verified against MC simulations using spectral numerical
	approximations. It is shown that the CDF model accurately predicts the
	mean and standard deviation for Gaussian, normal and beta distributions
	of the random coefficients.
\end{abstract}





\section{Introduction}
Deterministic predictions of models that couple multiphysics through source terms, such as chemically reactive, multiphase and/or multi-material flows, are notoriously difficult. A typical system involves many dependent variables, for which the exact coupling may not be known a priori. The coupling may also be imprecise because of measurement and/or numerical errors, and sparsity of experimental data. To improve upon the accuracy of the model, it is necessary to understand how uncertainty of these critical parameters propagates to a Quantity of Interest (QoI). Uncertainty quantification has become an integral component of computational models of chemically reacting flows (see \cite{Najm2009, Cai2016, Reagan2005, boso2018}, and the references contained therein) and particle-laden flow \cite{Fountoulakis2018, Shotorban2013, Guerra2016}. With a known uncertainty distribution in the parameter space, the areas for improvement of the dependent variables can be identified. 

If uncertainty is treated within the probabilistic framework, uncertain predictions of the QoI are fully captured by its probability density function (PDF) or cumulative ditribution function (CDF). The latter can be estimated via Monte Carlo (MC) simulations, but these converge slowly ($\sim 1/\sqrt{N_s}$, with $N_s$ the number of samples), and are hence computationally inefficient.  More efficient sampling techniques like multilevel MC \cite{Giles2008, Giles2013} and Latin hypercube sampling (LHS) \cite{Iman1980} have been developed, but may not alleviate computational cost. For example, the variance of LHS output samples can be larger than with normal sampling \cite{Owen2013}.

Generalized polynomial chaos (gPC) \cite{Wiener1938} provides a non-sampling alternative to MC. It expresses uncertain parameters in terms of orthogonal polynomials of standard random variables. These expansions can be used to obtain a spectral description of the uncertain parameters and are used in stochastic finite element methods (SFEMs) \cite{Stefanou2009}. SFEMs have been used to model a variety of phenomena, such as transport in porous media, solid mechanics, structural applications and reacting flow (see \cite{Reagan2005} for a review). The multi-element generalized polynomial chaos method \cite{Wan2005, Wan2006} can handle discontinuities in the stochastic space. Unfortunately, intrusiveness is a significant downside, i.e., a standard  numerical method requires modifications for gPC that result in a high-dimensional coupled linearized system of equations, which is computationally taxing \cite{Debusschere2017}. In fact, SFEM is computationally more expensive as compared to MC when a large number of random variables are considered.

Stochastic collocation methods (see \cite{Xiu2016} for an overview) are non-intrusive and require only a few repetitive calls to a deterministic solver, similar to MC. Adaptive sparse grid collocation \cite{Ma2009} handles discontinuities in the stochastic space. 
However, stochastic collocation can be slower than MC, e.g., for nonlinear parabolic equations with random coefficients with high variances \cite{Barajas2016}.

Statistical moment equations and the method of distributions provide yet another alternative to sampling methods. These approaches derive deterministic equations for statistical moments (typically, mean and variance) or PDF/CDF of a system state, respectively. Moment equations are derived through ensemble averaging, but they require a closure approximation. The closure is typically done through perturbation expansions or gradient models, which require empiricism and/or homogenization of higher fidelity data.
Either way, major concessions are made to model accuracy through closure. 
Equally important is the inability of the method of moments to deal with highly non-Gaussian system states, whose dynamics cannot be fully captured with their mean and variance. 
The method of distributions provides the full joint PDF/CDF of solution and the random inputs. PDF methods were first developed for applications in turbulence and combustion (see \cite{Pope2000}, for example, for a review), but later extended to quantify parametric uncertainty in a variety of problems  \cite{lichtner-2003-stochastic, tartakovsky-2009-probability, broyda-2010-probability, dentz-2010-probability, wang-2013-probability}. Crucially, PDF methods obviate the need for linearization of nonlinear terms. A drawback is the challenging definition of unique boundary conditions in stochastic space. 

We develop a CDF method for systems with stochastic sources as they may  appear in  multi-physics environments such as particle-laden flow and chemically reacting flow. By way of  example we consider a Burgers' equation with random source, which renders the system stochastic. An equation for the joint CDF of the flow velocity and source coefficients is derived. The marginal PDF for the uncertain velocity can then be extracted from this joint CDF. The main advantages of this method are its simplicity, accuracy and computational efficiency. Moreover, this method leads to an unambiguous closed system of equations. A simplified version of the CDF equation is solved using spectral methods, assuming the source consists of only one random coefficient and a steady smooth source. 
Solutions to the simplified CDF equation are shown to be in good agreement with MC results.

The paper is structured as follows. Section~\ref{sec:meth} describes the governing equations and the numerical methods. Special attention is given to the regularization of (singular) source terms that appear in problems with deterministic initial conditions. In Section 3, results are shown and discussed for the simplified CDF equation, assuming a uniform, normal or beta distribution for the random source coefficient. The influence of relevant parameters (e.g., grid resolution) is considered, and a thorough comparison is made with MC results. Conclusions and directions for future work are given in Section 4. 




\section{Governing equations and methodology}
\label{sec:meth}

\subsection{Burgers equations with a stochastic source}
Let a state variable/velocity $v(x,t)$, defined on $(x,t)\in[x_{\text{min}},x_{\text{max}}]\times\bR^+$, satisfy an inviscid Burgers equation with stochastic source,
\begin{equation}\label{eq:ProblemFormulation_1}
    \pd[v]{t}+v\pd[v]{x}=g^s(u-v)\cdot(u-v).
\end{equation}
The source term accounts for the relative velocity difference of the state variable $v$ and a (deterministic or random) background velocity $u$; and the functional form of $g^s(\cdot)$ is unknown/uncertain. Such a formulation is common in coupled multiphysics systems, where $u$ and $v$ represent the solution for two coupled physics environments, respectively (e.g., particle/gas flow, chemistry/gas flow).

Equation~\eqref{eq:ProblemFormulation_1} is subject to a deterministic initial condition $v(x,0)=v_{\text{in}}(x)$ and a deterministic boundary condition $v(x_{\text{min}},t)=v_0(t)$. The unknown (random) function $g^s(\cdot)$ is represented via a polynomial with orthogonal basis functions $\ChebT{i}(\cdot)$,
\begin{equation}\label{eq:ProblemFormulation_2}
    g^s(\cdot)=\sum_{i=0}^{\infty}a_i\ChebT{i}(\cdot)\approx \sum_{i=0}^{N_g}a_i\ChebT{i}(\cdot).
\end{equation}
We assume that $g^s$ has a compact support and choose $\ChebT{i}(\cdot)$ to be Chebyshev polynomials of the first kind, scaled from the interval $[-1,1]$ to $[x_{\text{min}},x_{\text{max}}]$. The polynomial coefficients are uncertain and treated as (correlated or not) random variables with the (joint) PDF $f_\va(A_1,\ldots,A_{N_g})$. Combining (\ref{eq:ProblemFormulation_1}) and (\ref{eq:ProblemFormulation_2}) yields a PDE with random coefficients,
\begin{equation}\label{eq:ProblemFormulation_3}
    \pd[v]{t}+v\pd[v]{x}=(u-v)\sum_{i=0}^{N_g}a_i\ChebT{i}(u-v),
\end{equation}
whose solution is given in terms of $f_v(V;x,t)$, the PDF of the random state variable $v(x,t)$. Equation (\ref{eq:ProblemFormulation_3}) can be solved with MC simulations, i.e., by repeatedly sampling the random coefficients and solving the corresponding deterministic PDEs. 

\subsection{CDF Equations}
\subsubsection{Positive source}
In Appendix A we show that the joint CDF $F_{\mathbf a v}(\mathbf A,V;x,t)$ of the set of input parameters $\mathbf a = \{a_0,\ldots,a_{N_g}\}$ and the state variable $v$ at any space-time point $(x,t)$ satisfies a deterministic integro-differential equation
\begin{equation}\label{eq:jCDF}
 \frac{\partial
	F_{\mathbf a v}}
	{\partial t} + V \frac{\partial F_{\mathbf a v}}{\partial x} = - (u-V) \sum_{i=0}^{N_g} T_i(u-V) \frac{\partial }{\partial V} \left[  A_i F_{\mathbf a v} - \int\limits_{-\infty}^{A_i} F_{\mathbf a v} (\mathbf A \!\setminus\! A_i, A_i',V;x,t) \text d A_i' \right],
\end{equation}
where
\begin{equation}
    \vA\!\setminus\! A_i = \left(A_{1},\ldots,A_{i-1},A_{i+1},\ldots,A_{N_g}\right).
\end{equation}
The CDF of random velocity $v$ at point $(x,t)$, i.e., $F_v(V;x,t)$, is the marginal of $F_{\mathbf a v}$, 
\be
    F_v(V;x,t) = F_{\mathbf a v}(\mathbf A_{\max},V;x,t).
\ee
Likewise,
\be
    F_{a_i v}(A_i',V;x,t) = F_{\mathbf a v}(\mathbf A_{\max}\!\setminus\! A_i,A_i',V;x,t),
\ee
Hence, the marginal $F_v$ satisfies a CDF equation
\begin{equation}\label{eq:CDF}
 \frac{\partial F_v}{\partial t} + V \frac{\partial F_v}{\partial x} =
 - (u-V) \sum_{i=0}^{N_g} T_i(u-V) \frac{\partial }{\partial V} \left[  A_{\max, i} F_v - \int\limits_{-\infty}^{A_{\max, i}} F_{a_i v} (A_i',V;x,t) \text d A_i' \right].
\end{equation}
However, we found it more convenient to solve the equation for the joint CDF. For now, we assume the initial velocity to be deterministic. This leads to the following initial conditions
\begin{align}\label{eq:CDFeq_IC}
F_{a_i v} (A_i,V;x,0) =  F_{a_i} (A_i) F_{v} (V;x,0) = F_{a_i} (A_i) \mathcal H(V - v_\text{in}(x)),\quad i=0,\ldots,N_g,
\end{align}
where $\mathcal H$ denotes the Heaviside function. Basic properties of probability yield boundary conditions
\begin{align}
F_{a_i v} (A_{\min,i},V;x,t) = 0,& \qquad  F_{a_i v} (A_{\max,i},V;x,t) = F_v(V;x,t), \nonumber\\
F_{a_i v} (A_i,V_{\min};x,t) = 0,& \qquad  F_{a_i v} (A_i,V_{\max};x,t) = F_{a_i}(A_i),
\end{align}
where  $i=0,\ldots,N_g$.

\subsubsection{Negative source}

While the CDF equation is applicable for arbitrary smooth sources with compact support, the numerical method described in the next section turns out to be unstable for negative sources. This is likely due to undershoots leading to negative CDFs and steepening of the CDF. 
%
As a workaround, we solve the CDF equation 
\begin{equation}\label{eq:Remarks_CDFequation_NegativeSource}
    \frac{\partial
	G_{a v}}
	{\partial t} + V \frac{\partial G_{a v}}{\partial x} = - \left(u-V\right)\frac{\partial }{\partial V} \left[  A G_{a v} - \int\limits_{-\infty}^{A} G_{a v} (A',V;x,t) \text d A' \right],
\end{equation}
where
\begin{equation}
    G_{a v} (A,V;x,t)=F_a(A)-F_{a v} (A,V;x,t).
\end{equation}
Following the same steps as in Appendix A, but with $\tilde{\Pi}(A,a;V,v) = \mathcal H(A-a)- \Pi(A,a;V,v)$, one can prove that the function $G$, which is not a CDF, does satisfy the CDF equation.  It is subject to the following adjusted initial and boundary conditions 
\begin{equation}\label{eq:Remarks_CDFequation_NegativeSource_IC}
    G_{a v} (A,V;x,0) =  F_a(A)\left(1-\mathcal H(V-1)\right) =F_{a} (A) \mathcal H(1-V),
\end{equation}
and
\begin{align}
    G_{a v} (A_{\min},V;x,t) &= 0,& \qquad  G_{a v} (A_{\max},V;x,t) &= 1-F_v(V;x,t), \nonumber\\
    G_{a v} (A,V_{\min};x,t) &= F_{a}(A),& \qquad  G_{a v} (A,V_{\max};x,t) &= 0.\label{eq:Remarks_CDFequation_NegativeSource_BC}
\end{align}
While $G_{av}$ is not a CDF; it is, for example, no longer non-decreasing in each of its variables, the corresponding PDF can be determined as follows
\begin{equation}
    f_v(V;x,t)=-\frac{\partial^2}{\partial A \partial V} G_{av}(A_{\max},V;x,t).
\end{equation}
For the sake of simplicity, we consider one random coefficient $a$, instead of a random vector $\va$. The extension of the result to multiple coefficients is straightforward.

\subsection{Numerical Methods}

Numerical solutions of the CDF equation are tested against solutions of Equation (\ref{eq:ProblemFormulation_3}) obtained by MC simulations. 
Both Equation (\ref{eq:ProblemFormulation_3}) and (\ref{eq:jCDF}) admit solutions with singularities for several reasons. First, the deterministic initial condition~(\ref{eq:CDFeq_IC}) contains a Heaviside function. Second, the Burger's equation is known to steepen solutions leading to discontinuities. Finally, certain reduced-physics models can have singular sources \cite{Suarez2014}. To obtain accurate solutions and consistency between MC and the CDF equation, we must be careful in selecting the numerical methods that we use to approximate the governing systems. Here, we rely on a low dispersive and low diffusive, single domain Chebyshev collocation method and use some of the recently developed filtering and regularization techniques to capture shocks and regularize sources \cite{Suarez2014,Wissink2018}. In the following, we briefly summarize the Chebyshev collocation method and the regularization techniques. For a detailed discussion the interested reader is referred to~\cite{Gottlieb2001,Suarez2014,Wissink2018,Hesthaven}.

\subsubsection{Chebyshev Collocation Method and Time Integration}

The collocation method is based on polynomial interpolation of a function $u(x)$, and can be expressed as
\begin{equation}
  u_{N_x}(x) = \sum\limits_{j=0}^{N_x} u(x_j) l_j(x), \quad l_j(x)=\prod_{k=0,k\neq j}^{N_x} \frac{x-x_k}{x_j-x_k} \qquad ,\quad j = 0, \ldots,N_x,
\label{eq:interpolant}
\end{equation}
where $x_j$ with $j = 0,\ldots,N_x$ are collocation points, and $l_j(x)$ are the Lagrange interpolation polynomials of degree $N_x$.
To determine the derivative of the function $u(x)$ at the collocation points $x_i$, $u'(x_i)$, one can simply take the derivative of the Lagrange interpolating polynomial as
\begin{equation}
\frac{\partial u(x_i)}{\partial x} \approx \sum\limits_{j=0}^{N_x} u(x_j) l_j'(x_i),
\end{equation}
or, written compactly in the matrix-vector multiplication form as
\begin{equation}
{\vec u}' = \vD \vec u,
\label{eq:spatialdisc}
\end{equation}
where the differentiation matrix $D_{i,j}=l_j'(x_i)$.
For the Chebyshev collocation method, the collocation points are chosen at the Gauss-Lobatto quadrature points,
\begin{equation}
  \xi_i=-\cos(i\pi/N_x), \quad i=0,\ldots,N_x,
\end{equation}
such that the $L_\infty$ norm of the interpolant is minimized on its interval [-1,1]. Combining Equation (\ref{eq:ProblemFormulation_3}) and (\ref{eq:spatialdisc}) results in a system of ordinary differential equations (ODEs) on the collocation points $\boldsymbol{x}\subseteq [x_{\mathrm{min}},x_{\mathrm{max}}]$:
\be
    \der[\vv(t)]{t} = \text{diag}(\vu-\vv(t))\sum_{k=0}^{N_g}a_k\ChebT{k}(\vu-\vv(t))-\text{diag}(\vv(t))\vD_x\vv(t),
\ee
in which
\begin{align*}
  &\vu = [u(x_0),\ldots,u(x_{N_x})]^T,\quad \vv(t) = [v(x_0,t),\ldots,v(x_{N_x},t)]^T,\\  &\ChebT{k}(\vu-\vv(t))= [\ChebT{k}((u(x_0)-v(x_0,t)),\ldots,\ChebT{k}((u(x_{N_x})-v(x_{N_x},t))]^T,
\end{align*}
diag$(\vx)$ denotes a diagonal matrix with entries $x_i$, $i=0,\ldots,N_x$, and $\vD_x={\partial{\xi}/\partial{x}}\times\vD=2/(x_{\mathrm{max}}-x_{\mathrm{min}})\times\vD$ is a scaled version of $\vD$ to account for the mapping of the spatial domain from the Chebyshev quadrature nodes $\xi$ to the spatial domain $\vx$. To integrate the system of ODE's in time, we employ a fourth order Runge-Kutta scheme \cite{Press1992} for the MC-equation.
Using orthogonality of the Chebyshev polynomial, the CDF equation is similarly discretized on a tensorial Gauss-Lobatto grid in ($x$,$V$), and given for every point $\tilde{\vA}$ on the tensorial uniform $\vA$-grid$\subseteq [A_{\mathrm{min}},A_{\mathrm{max}}]^{N_g}$ by
\begin{align}\label{eq:discrete_CDF}
    \der[\vF_{\va v}]{t}(t) &= - \vD_x \vF_{\va v}(t) \text{diag}(\vV)\nonumber\\
     &\hspace{1.5cm}- \sum_{k=0}^{N_g} \vD_V \left[  \tilde{A}_i \vF_{\mathbf a v}(t) - \int\limits_{-\infty}^{\tilde{A}_i} \vF_{\mathbf a v} (\mathbf{\tilde{A}} \!\setminus\! \tilde{A}_i, \tilde{A}_i',t) \text d \tilde{A}_i' \right] \text{diag}\left[\text{diag}(\vu-\vV)\ChebT{k}(\vu-\vV)\right],
\end{align}
 with $\vV = [V_0,\ldots,V_{N_V}]^T$ the grid along the $V$-direction and $\vD_V=2/(V_{\mathrm{max}}-V_{\mathrm{min}})\times\vD$ a scaled differentiation matrix.
$\vF_{\va v}(t)$ is a $(N_x+1)\times (N_V+1)$-matrix given by $\vF_{\va v}^{i,j}(t)=\vF_{\va v}(V_j,x_i,t)$. We have taken an equal amount of grid points in $x$- and $V$-direction, i.e., $N_x=N_V$. Because we use a tensorial grid, the integral on the RHS of Equation~\eqref{eq:discrete_CDF} can be simply evaluated along lines in the $A$-coordinate direction. We found that the trapezoid rule was sufficiently accurate.

Because of inherent sharp gradients in the solution of the CDF equations, a third order Total Variation Diminishing (TVD) Runge-Kutta scheme \cite{Gottlieb1998}
\begin{eqnarray}\label{eq:RK_TVD}
  u^{(1)}&=&u^n+\Delta t L(u^n),  \nonumber \\
  u^{(2)}&=&\frac{3}{4}u^n+\frac{1}{4}u^{(1)}+\frac{1}{4}\Delta t L(u^{(1)}), \\
  u^{n+1}&=&\frac{1}{3}u^n+\frac{2}{3}u^{(2)}+\frac{2}{3}\Delta t L(u^{(2)}), \nonumber
\end{eqnarray}
is used to reduce numerical oscillations induced in time. In  Equation~\eqref{eq:RK_TVD},  $L$ denotes the discrete spatial derivative operator. 

\subsubsection{Filtering}

Following  \cite{Wissink2018}, we filter the solution as follows
\be\label{eq:temp1}
    \tilde{u}(x)=\int_{x-\epsilon}^{x+\epsilon}u(\tau)\delta_\epsilon^{m,k}(x-\tau)d\tau,
\ee
using a kernel that regularizes the Dirac delta function with a class of high-order, compactly supported piecewise polynomial \cite{Suarez2014},
\begin{equation}\label{eq:filtering_delta_pol_kernel}
    \delta_\epsilon^{m,k}(x)=\begin{cases}
                               \tfrac{1}{\epsilon}P^{m,k}(\tfrac{x}{\epsilon}), & x\in [-\epsilon,\epsilon] \\
                               0, & \mbox{otherwise},
                             \end{cases}
\end{equation}
where $\epsilon>0$ is the support width or scaling parameter. The polynomial $P^{m,k}$ controls the number of vanishing moments $m$, and the number of continuous derivatives at the endpoints of the compact support $k$. 
In \cite{Suarez2014} it was shown that the filter based on the Dirac-delta approximation $\delta_\epsilon^{m,k}$ converges according to $\order{\epsilon^{m+1}}$ in smooth solution regions away from regularization areas.
Filtering of the interpolant $u_N$ (\ref{eq:interpolant}) leads to
\begin{equation}
    \tilde{u}_{N_x}(x)=\int_{x-\epsilon}^{x+\epsilon}\left[\sum\limits_{i=0}^{N_x} u(x_i) l_i(\tau)\right]\delta_\epsilon^{m,k}(x-\tau)d\tau = \sum\limits_{i=0}^{N_x} u(x_i) S_i(x),
\end{equation}
after interchanging summation and integration, where the filtering function $S_i$ is given by
\begin{equation}
    S_i(x)=\int_{x-\epsilon}^{x+\epsilon} l_i(\tau)\delta_\epsilon^{m,k}(x-\tau)d\tau.
\end{equation}
On the discrete collocation points the convolution reduces to a matrix vector product
\begin{equation}
    \vec{\tilde{u}}=\vS \vec{u},
\end{equation}
where the $(N_x+1)\times(N_x+1)$ filtering matrix $\vS$ has the elements
\begin{equation}\label{eq:Filtering_Implementation_Filtermatrix}
    S_{i,j} = \int_{x_j-\epsilon}^{x_j+\epsilon} l_i(\tau)\delta_\epsilon^{m,k}(x_j-\tau)d\tau.
\end{equation}
The filtering matrix $\vS$ can be pre-computed. For the solution of the CDF equation, the filtering procedure is applied to regularize the Heaviside function in the initial joint CDF $F_{\va v}(A,V;x,t=0)=F_\va(A)\mathcal{H}(V-1)$ at every $A$-grid point.

\subsubsection{Exponential filter}
\begin{figure}
  \centering
  \includegraphics[width=\textwidth]{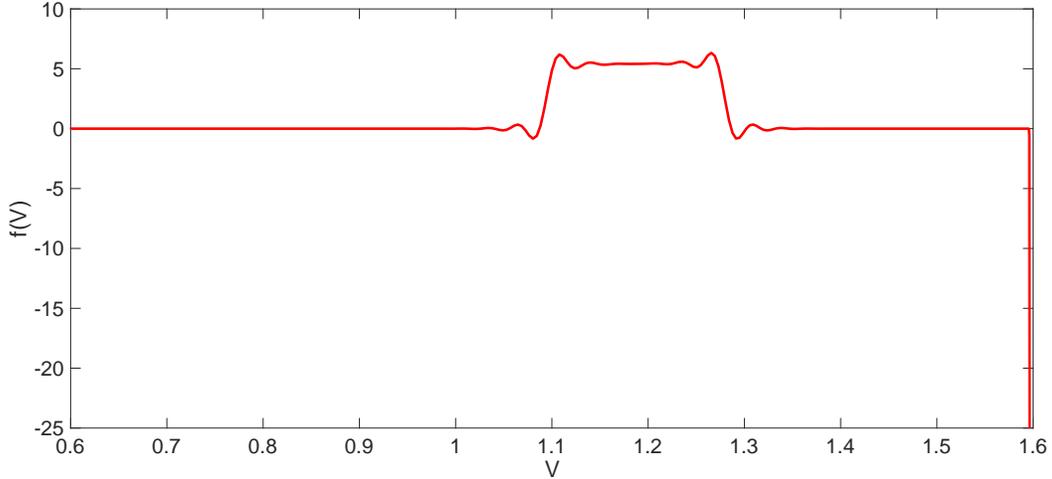}
  \caption{PDF $f_v(V)$ at $t=5\cdot 10^{-3}$ and $x=0.03$ obtained by solving the CDF equation (\ref{eq:setup_CDFequation}), assuming $a\sim\cU({[0.5,1.5]})$. The regularization filter corresponding to $k=8, m=13, N_d=50$ is used. Further, $N_x=400$ and $N_a=10$. No exponential filter has been applied.} \label{fig:filtering_ExampleGoodFilter}
\end{figure}

We found a sharp increase in the marginal ($V$-)CDF solution at $V=\vmax$ (Fig.~\ref{fig:filtering_ExampleGoodFilter}, for example). At this location no boundary condition is specified because the  solution moves out of the domain along its characteristic. The boundary spike does not always lead to instability and does not affect the meaningful CDF solution in the center of the domain. It is not entirely clear what the cause for this increase is, but the global nature of the collocation approximation and hence global sensitivity of the solution, can easily yield this type of change at the boundary. 
We  found that an exponential filter applied to the solution in the regions close to boundary where the CDF is near constant, suppressed the undesired boundary behavior. We use the filter according to \cite{Hesthaven,Vandeven1991} as follows:
\be
    \tilde{F}_{av} = S_{\text{exp}} F_{av},
\ee
where the elements of the $p$-th order $(N_x+1)\times (N_x+1)$ exponential filter matrix $S_{\text{exp}}$ are given by
\be
    S_{\text{exp}}^{i,j} = \frac{2}{c(j)N_x}\sum_{k=0}^{N_x} \frac{1}{c(k)}\left(e^{-\alpha (k/N_x)^{p}}\cos\left(\frac{i k\pi}{N_x}\right)\cos\left(\frac{j k\pi}{N_x}\right) \right),
\ee
in which $\vc$ is the $(N_x+1)$-dimensional vector
\be
    \vc = [2,1,1,\ldots,1,2]
\ee
and $\alpha=-\ln(10^{-16})$. In general $p$=5 suppressed the spike.

\subsubsection{Sampling the PDF}\label{subsec:MC-simulation}

To compare MC against the CDF model, we must determine a PDF from a set of samples. For this, we use the built-in Matlab kernel density estimator \mcode{ksdensity} \cite{ksdensity}. A kernel distribution is a nonparametric representation of the PDF of a random variable. It is used when a parametric distribution cannot properly describe the data, or when one wants to avoid making assumptions about the distribution of the data. A kernel distribution is defined by a smoothing kernel and a bandwidth value, which control the smoothness of the resulting density curve. We refer the interested reader to \cite{Silverman1986} for more information on density estimation. The kernel density estimator's formula is given by
\begin{equation}\label{eq_densityestimator}
    \hat{f}_{Bw}(\xi)=\frac{1}{N_s Bw}\sum_{i=1}^{N_s}K\left(\frac{\xi-\xi_i}{Bw}\right),
\end{equation}
where $\xi_1, \xi_2,\ldots, \xi_{N_s}$ are random samples from an unknown distribution, $N_s$ is the sample size, $K(\cdot)$ is the kernel smoothing function, and $Bw$ is the bandwidth.

Here, we choose the density estimate produced by \mcode{ksdensity} to be based on a normal (Gaussian) kernel function. Other kernels, notably the box, triangle or Epanechnikov kernel, can also be used. Details about the used parameter values are given in Section \ref{subsec:MC simulation}.

\subsection{Setup for Numerical Tests}
To verify consistency between MC and the CDF model, we consider only the first random coefficients in  $\va=\{a_0,\ldots,a_{N_g}\}$, i.e. $\va=a_0$.  Further, the source will be assumed to be a steady Gaussian source,
\begin{equation}
    u(x)=\tfrac{1}{\sqrt{2\pi}\sigma}e^{\frac{(x-x_a)^2}{2\sigma^2}},
\end{equation}
where $x_a$ denotes the center of the source and $\sigma$ is a measure for the width of the support. This results in the following Burger's test model
\begin{equation}\label{eq:setup_MCequation}
    \pd[v]{t}+v\pd[v]{x}=a\left(\tfrac{1}{\sqrt{2\pi}\sigma}e^{\frac{(x-x_a)^2}{2\sigma^2}}-v\right).
\end{equation}
on the spatial interval $x\in [0,0.06]$. The source is placed in the center of the interval, so $x_a=0.03$. Further, the spread of the source is given by $\sigma=5\cdot 10^{-3}$. 
We assume that the initial velocity $v$ is deterministic, satisfying boundary and initial conditions
\begin{equation}\label{eq:setup_MCbcandic}
    v(x,0)=1,\quad v(0,t)=1.
\end{equation}
The initial-boundary value problem (\ref{eq:setup_MCequation}-\ref{eq:setup_MCbcandic}) is solved using MC simulation.

The CDF-equation (\ref{eq:jCDF}) then takes the following form
\begin{equation}\label{eq:setup_CDFequation}
    \frac{\partial
	F_{a v}}
	{\partial t} + V \frac{\partial F_{a v}}{\partial x} = - \left(\tfrac{1}{\sqrt{2\pi}\sigma}e^{\frac{(x-x_a)^2}{2\sigma^2}}-V\right)\frac{\partial }{\partial V} \left[  A F_{a v} - \int\limits_{-\infty}^{A} F_{a v} (A',V;x,t) \text d A' \right]
\end{equation}
on $(x,V,A)\in [0,0.06]\times[V_{\min},V_{\max}]\times[A_{\min},A_{\max}]$, with initial condition
\begin{equation}\label{eq:setup_CDFequation_IC}
    F_{a v} (A,V;x,0) =  F_{a} (A) F_{v} (V;x,0) = F_{a} (A) \mathcal H(V - v_\text{in}(x)),
\end{equation}
with $v_\text{in}(x)=1$, and boundary conditions
\begin{align}
    F_{a v} (A_{\min},V;x,t) = 0,& \qquad  F_{a v} (A_{\max},V;x,t) = F_v(V;x,t), \nonumber\\
    F_{a v} (A,V_{\min};x,t) = 0,& \qquad  F_{a v} (A,V_{\max};x,t) = F_{a}(A).\label{eq:setup_CDFequation_BC}
\end{align}
We take $V_{\min}=0.6$, $V_{\max}=1.6$, $A_{\min}=0.5$ and $A_{\max}=1.5$.
To test consistency for a number of CDF distributions, we consider a uniform ($a\sim\cU({[0.5,1.5]})$), normal ($a\sim\cN(1,0.15)$) and beta distribution ($a\sim\cB(2,5)+0.5$) for the random source coefficient $a$, all with their density concentrated in the interval $[0.5,1.5]$. The beta distribution is translated by 0.5 to the right, to ensure that the corresponding random variable mainly takes values in $[0.5,1.5]$, but to improve readability we will denote it by $a\sim\cB(2,5)$ in what follows. By taking this set of distributions, we test the CDF method on discontinuous, smooth and skewed density functions. 

Following numerical experiments (see next chapter), we take $N_V$=100 grid intervals (hence 101 grid points) in the $V$-direction and $N_s$=20,000 samples of the random coefficient $a$ to solve the MC equation~\eqref{eq:setup_MCequation}, while in the CDF-routine we consider $N_x$=$N_V$=400 grid intervals in the $x-$ and $V-$ direction, and $N_A$=10 grid intervals in the $A$-direction to solve Equation~\eqref{eq:setup_CDFequation}. Further, use a regularization filter corresponding to $k=8, m=13$, and $N_d=50$, which gives good (stable and filter-independent) results up to 400 intervals in $V$-direction for the CDF equation. The filter is applied twice (to ensure sufficient smoothness of the Heavisde) to the initial joint CDF $F_{\va v}(A,V;x,t=0)=F_\va(A)\mathcal{H}(V-1)$ at every $A-$grid point. Regularization is not needed in the MC-routine, since all functions and solutions are smooth there. In the CDF routine, an exponential filter of order $p$=5 is applied after every time step to the 50 rightmost $V-$grid points (see next chapter). The integral in Equation (\ref{eq:setup_CDFequation}) is approximated using the trapezoidal rule.



\section{Results}
In this section we test consistency between MC and the CDF model. First, the effect of different parameter values (grid resolution, sample size, etc.) are considered for both MC and CDF. Subsequently, the MC-solution with the highest number of samples and optimal bandwidth and CDF-solution with the finest grid are directly compared. 
All results are obtained with Matlab 2018a. 

\subsection{Monte-Carlo simulation}\label{subsec:MC simulation}
In the MC simulation, the accuracy of the solution depends on a number of parameters, including the number of spatial grid intervals $N_x$, the number of samples $N_S$, time step $\delta t$ and the input parameters of the density estimator (type of smoothing kernel, number of grid points $N_{\mathrm{ks}}$ and bandwidth). Below we discuss the effect of each of these parameters on the solution.

\begin{figure}
	\centering
    \includegraphics[width=\textwidth]{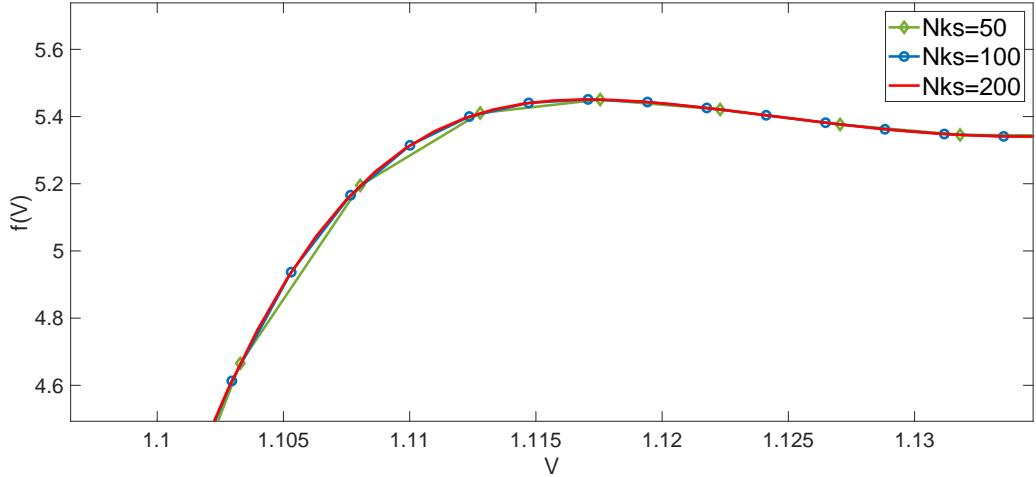}
    \caption{Closeup of PDF $f_v(V)$ at $t=5\cdot 10^{-3}$ and $x=0.03$ obtained by MC simulation, assuming $a\sim\cU({[0.5,1.5]})$. In this figure, $N_x=100$, $N_S$=20,000 and $N_{\mathrm{ks}}$=50, 100 or 200. The optimal bandwidth is used.}
    \label{fig:Results_MC_Nksdensity_Compare}
\end{figure}

\begin{figure}
	\centering
    \includegraphics[width=\textwidth]{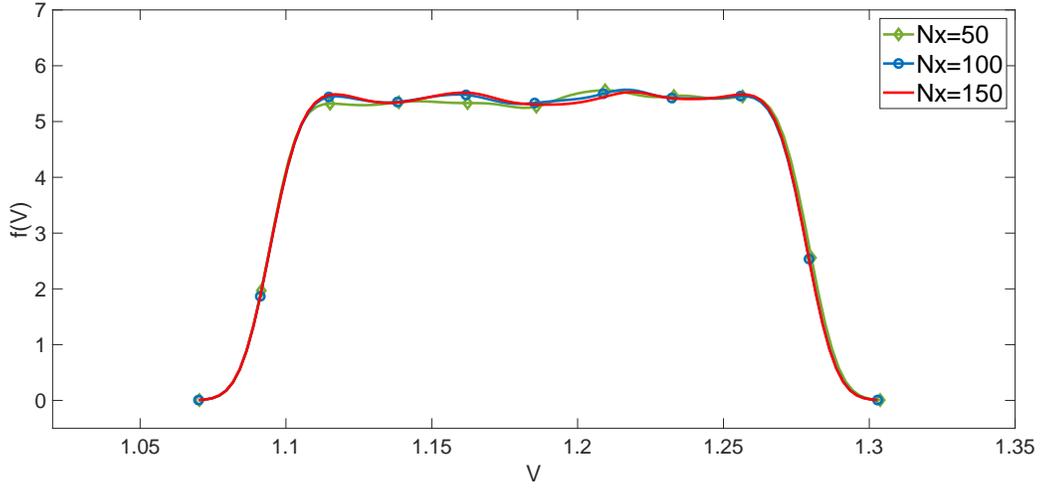}
    \caption{PDF $f_v(V)$ at $t=5\cdot 10^{-3}$ and $x=0.03$ obtained by MC simulation, assuming $a\sim\cU({[0.5,1.5]})$. In this figure, $N_S$=20,000, $N_{\mathrm{ks}}$=100 and $N_x$=50, 100 or 150. The optimal bandwidth is used.}
    \label{fig:Results_MC_Nx_Compare}
\end{figure}

Figure \ref{fig:Results_MC_Nksdensity_Compare} shows the effect of the number of grid points $N_{\mathrm{ks}}$ in $V-$direction in the density estimation, with $a\sim\cU({[0.5,1.5]})$, $N_x$=100, $N_S$=20,000 and the optimal bandwidth (see below). It is clear that the default number of grid points $N_{\mathrm{ks}}$=100 provides sufficient accuracy. Likewise, Figure \ref{fig:Results_MC_Nx_Compare} depicts the effect of $x-$grid resolution, with $a\sim\cU({[0.5,1.5]})$, $N_x$=50, 100 or 150, $N_S$=20,000 and the optimal bandwidth. We conclude that $N_x$=100 gives sufficiently accurate results.
Therefore, in all MC-results $N_x$=100 intervals and $N_{\mathrm{ks}}$=100 are considered. The time step $\Delta t$ is a function of the number of spatial grid points, to ensure stability. It is given by $\Delta t=\lambda(x_{\max}-x_{\min})/(N_x+1)^2$, where $\lambda=1.5$ is the CFL condition number.

\begin{figure}
	\centering
    \includegraphics[width=\textwidth]{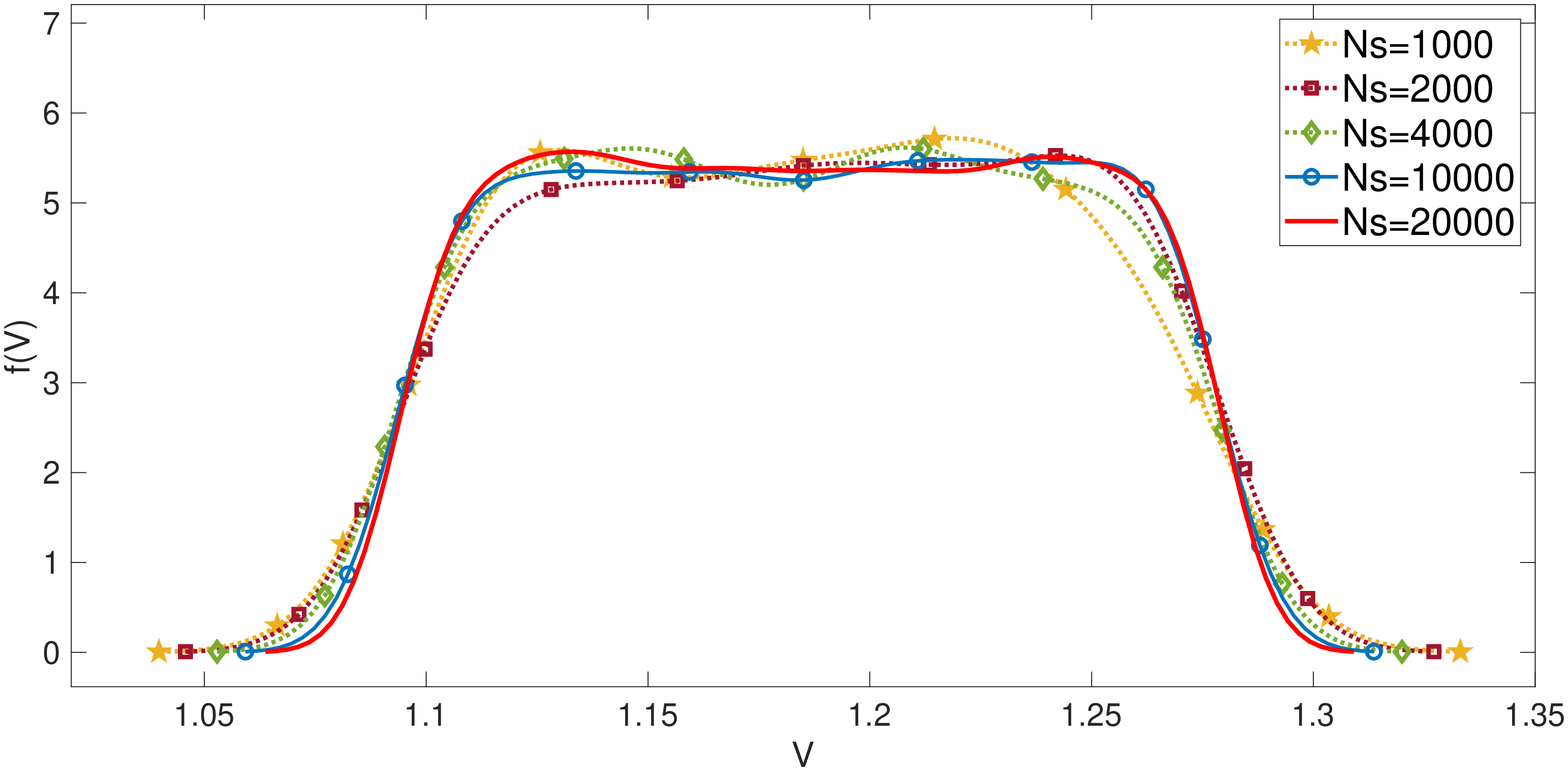}
    \caption{PDF $f_v(V)$ at $t=5\cdot 10^{-3}$ and $x=0.03$ obtained by MC simulation, assuming $a\sim\cU({[0.5,1.5]})$. In this figure, $N_x=100$ and the number of samples $N_S$ is either $1000, 2000, 4000, 10,000$ or $20,000$. The default bandwidth is used.}
    \label{fig:Results_MC_samplesize}
\end{figure}

\begin{figure}
	\centering
    \includegraphics[width=\textwidth]{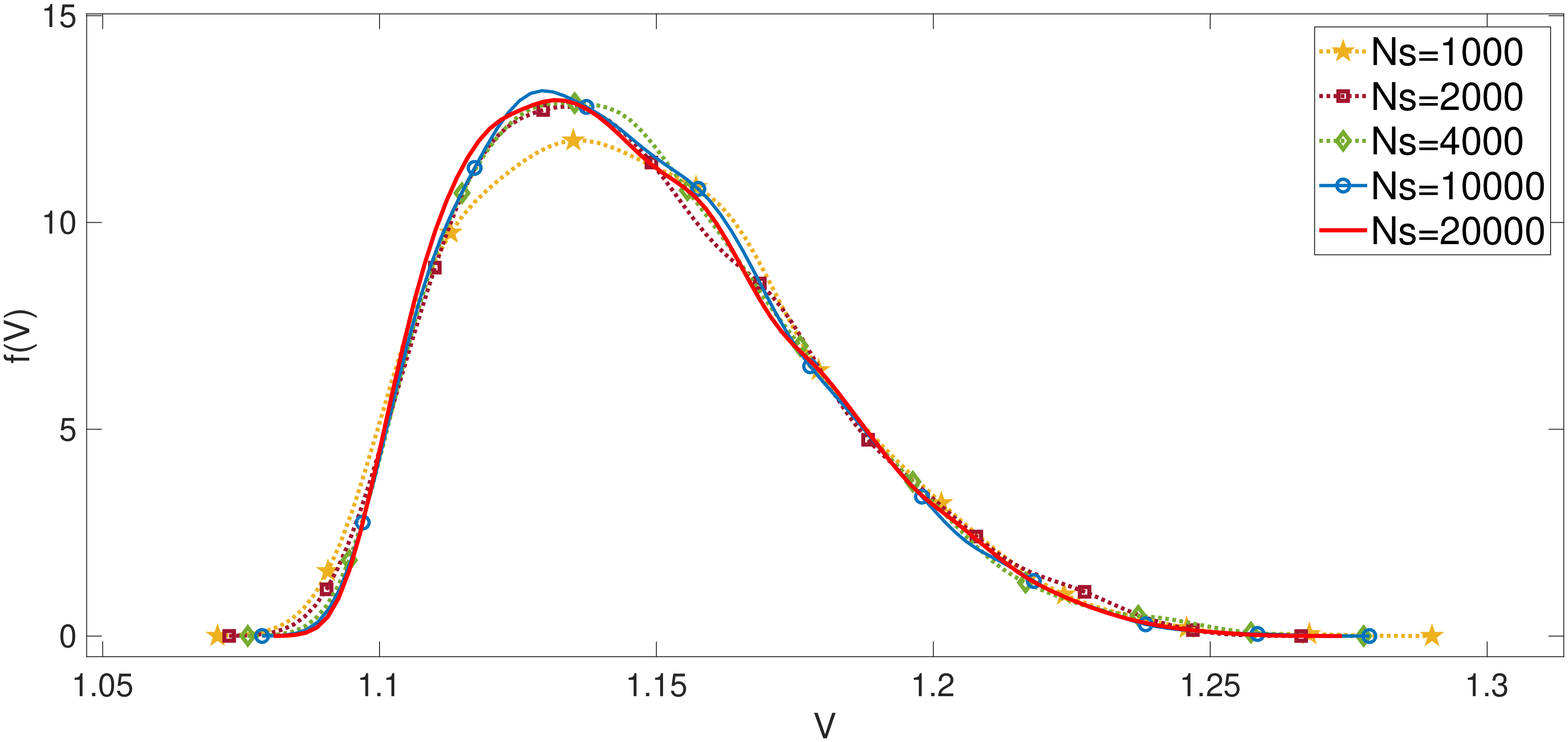}
    \caption{PDF $f_v(V)$ at $t=5\cdot 10^{-3}$ and $x=0.03$ obtained by MC simulation, assuming $a\sim\cB(2,5)$. In this figure, $N_x=100$ and the number of samples $N_S$ is either $1000, 2000, 4000, 10,000$ or $20,000$. The default bandwidth is used.}
    \label{fig:Results_MC_samplesize_Beta}
\end{figure}

Figure \ref{fig:Results_MC_samplesize} shows the effect of sample size for the problem with a uniform distribution. Likewise, Figure \ref{fig:Results_MC_samplesize_Beta} shows results for a beta distributed random coefficient. For $N_S$=20,000, we deem the density function to be sufficiently accurately resolved and consider it for further comparison in the paper. A further increase of the number of samples leads to a computational cost that is not feasible within the computational resources available to us.

\begin{figure}
	\centering
    \includegraphics[width=\textwidth]{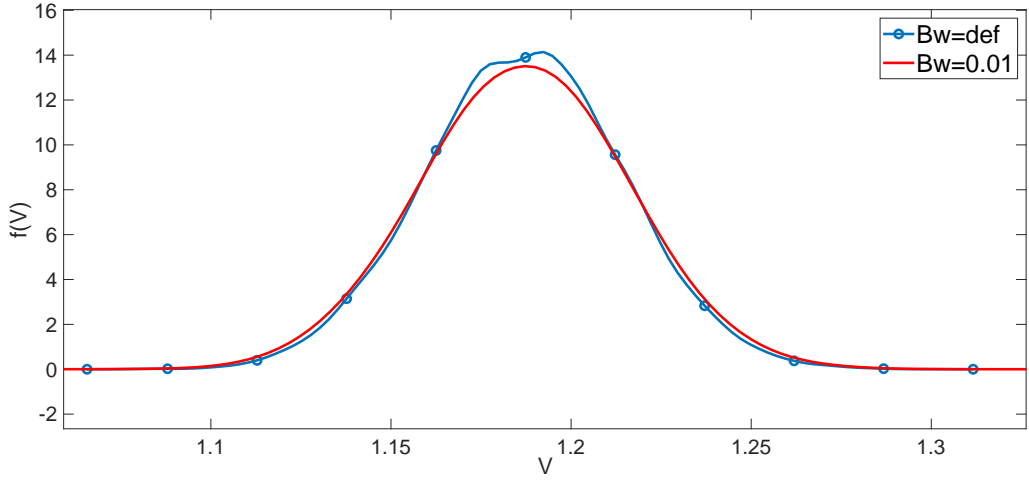}
    \caption{PDF $f_v(V)$ at $t=5\cdot 10^{-3}$ and $x=0.03$ obtained by MC simulation, assuming $a\sim\cN(1,0.15)$. In this figure, $N_x$=100, $N_s$=20,000 and both the default and optimal bandwidth are shown.}
    \label{fig:Results_MC_Bandwidths}
\end{figure}

Lastly, a remark about the density estimator. The bandwidth has a significant impact on the accuracy of the PDF determination. If it is taken too large, there is much overlap between the individual kernels (see Eq. (\ref{eq_densityestimator})). Then with the density estimate being the sum of almost identical Gaussians, the PDF will resemble a normal distribution, thereby obscuring the underlying behavior. If the bandwidth is taken too small, the support of most individual kernels - especially those belonging to sample outliers - will be isolated, causing the estimate to look like a sum of independent Gaussians with small support, i.e., with multiple steep peaks. In other words, sample outliers are not ignored in this case, but influence the density estimate significantly. To amend this issue, we empirically determine an optimal bandwidth for each distribution, by taking the smallest bandwidth that yields a PDF resembling that of a single random variable. For the uniform distribution, this is not possible. In this case, a bandwidth is chosen such that the support of the PDF resembled the support of the actual sample, while avoiding too much peaks in the center part. Figure \ref{fig:Results_MC_Bandwidths} shows the difference in PDF between the default and optimal bandwidth for $a\sim\cN(1,0.15)$. It can be seen that the default bandwidth is too small, resulting in a multi-peaked distribution. By carefully increasing the bandwidth, one obtains a single-peaked distribution, without losing the underlying behaviour - the location of the maximum and the tail behaviour remains very similar. The so-called optimal bandwidths are $0.008$, $0.01$ and $0.01$ for uniform, normal and beta-distributed $a$, respectively.
\subsection{CDF equation}
For the CDF equation, the solution accuracy is affected by the grid resolution in $A$-, $x-$ and $V-$ direction, as well as the regularization and exponential filter settings. Following \cite{Wissink2018} and discussed above, we choose the regularization filter with $m=13, k=8, N_d=50$. An exponential filter of order $p$=5 is applied locally, i.e., to the 50 rightmost $V$-grid points, to suppress oscillations near the boundary without impacting the solution at the center of the $V$-domain. Note that filtering locally is only possible because the solution is already approximately zero on the $51^{\text{st}}$ grid point (see Fig. \ref{fig:filtering_ExampleGoodFilter}, for example). The transition from non-filtered to filtered is thus smooth. This is generally not the case; one should then filter globally, which leads to smearing of the solution. A finer (higher order) filter only partly mitigates this issue.

\begin{figure}
	\centering
    \includegraphics[width=\textwidth]{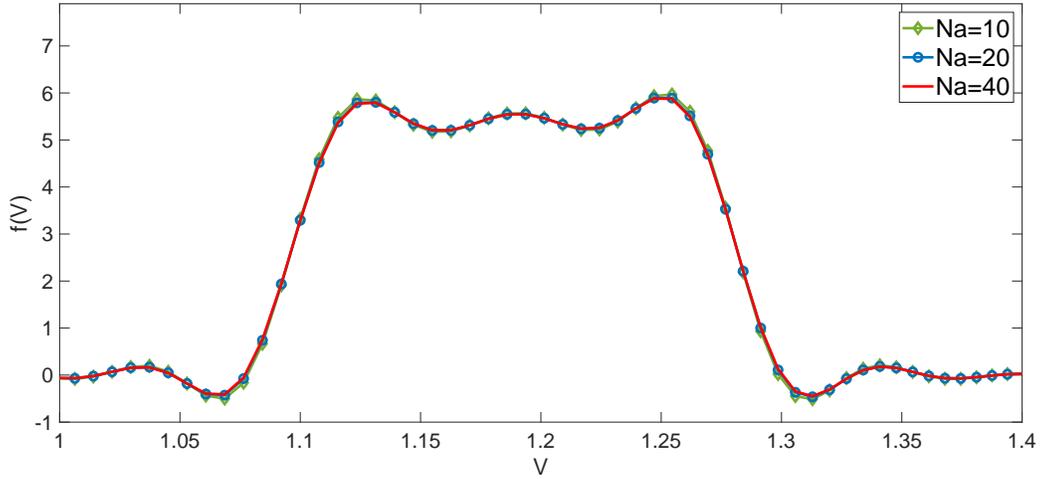}
    \caption{PDF $f_v(V)$ at $t=5\cdot 10^{-3}$ and $x=0.03$ obtained by solving the CDF-equation (\ref{eq:setup_CDFequation}), assuming $a\sim\cU({[0.5,1.5]})$. The regularization filter corresponding to $k=8, m=13, N_d=50$ is used. Further, $N_x$=$N_v$=200 and $N_a$=10, 20 or 40. A fifth order exponential filter has been applied to the 50 rightmost grid points.}
    \label{fig:Results_CDF_gridsizeA}
\end{figure}

\begin{figure}
	\centering
    \includegraphics[width=\textwidth]{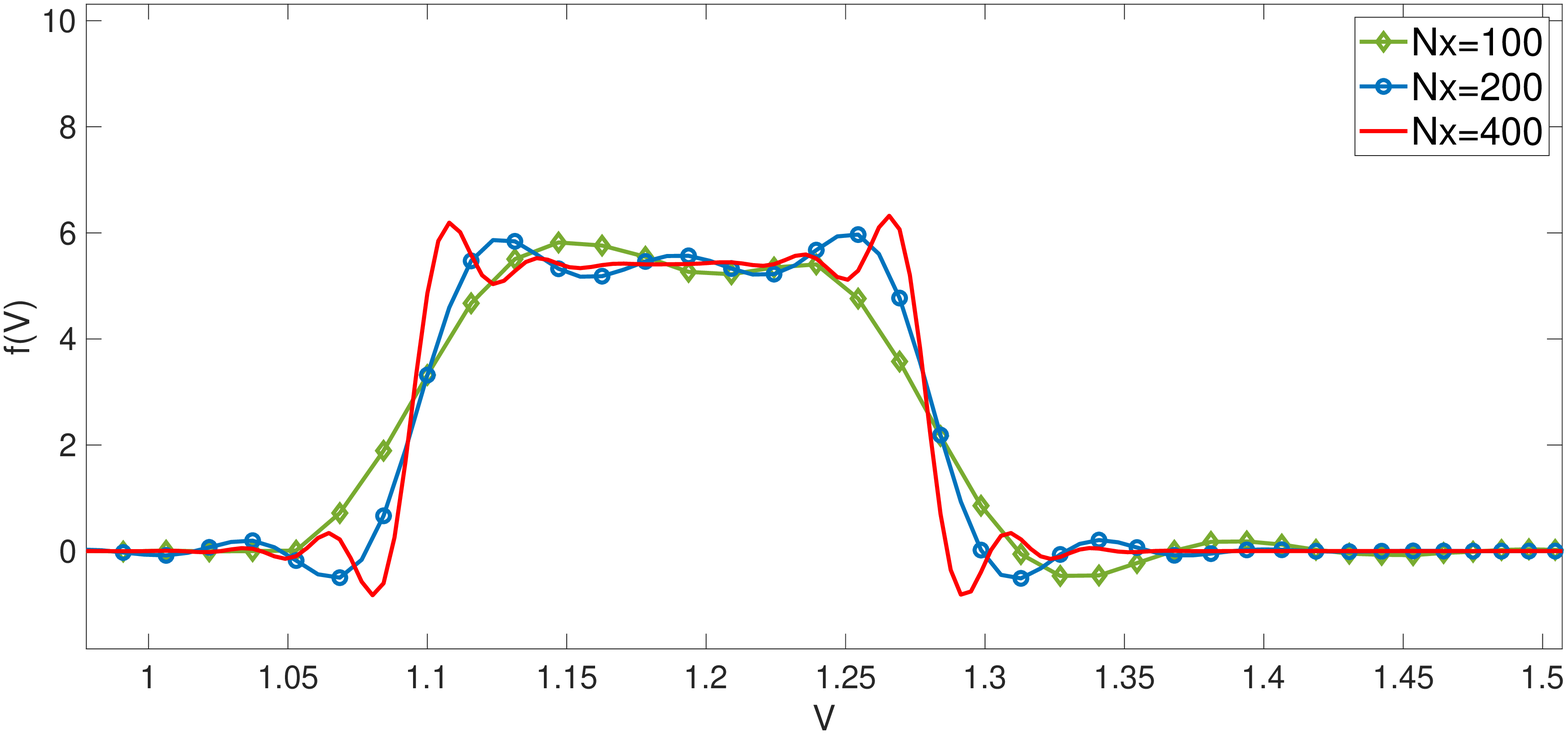}
    \caption{PDF $f_v(V)$ at $t=5\cdot 10^{-3}$ and $x=0.03$ obtained by solving the CDF-equation (\ref{eq:setup_CDFequation}), assuming $a\sim\cU({[0.5,1.5]})$. The regularization filter corresponding to $k=8, m=13, N_d=50$ is used. Further, $N_x$ and $N_v$ are either $100, 200$ or $400$ and $N_a=10$. A fifth order exponential filter has been applied to the 50 rightmost grid points.}
    \label{fig:Results_CDF_gridsize}
\end{figure}

In $A$-direction, as shown in Figure \ref{fig:Results_CDF_gridsizeA}, a number of grid intervals of $N_A$=10 converges the solution within the eyeball norm. 
In the $V$-direction the grid resolution has a significant effect. First of all, with increased resolution the regularization zone of possible discontinuities in the solution (CDF or PDF) can naturally be smaller as shown in Figure \ref{fig:Results_CDF_gridsize}.
In the reduced zone, larger under- and overshoot are visible that are induced by the non-positive delta kernel. These undershoots/overshoots are local and necessary to obtain high order convergence away from the regularization zone (see \cite{Wissink2018} for details). They are advected and visible at later time as seen in Figure \ref{fig:Results_CDF_gridsize}, for example. Secondly, as the grid resolution increases, the solution converges outside the regularization zone according to the theoretical convergence rate. For $N_x$ and $N_V$=400, the solution has converged, as shown in Figure \ref{fig:Results_CDF_gridsize}. The time step $\Delta t$ is given by $\Delta t=2\lambda(x_{\max}-x_{\min})/(V_{\max}N_x^2)$, where $\lambda=1.2$ is the CFL condition number.

\subsection{Comparison MC and CDF: positive source}
With grid-independence established for the  MC and CDF solution, we can now test consistency between the two methods. To do so, as a first indicator, we compare statistical moments - the mean, standard deviation, skewness and the zeroth moment at $x=0.03$. Secondly, a visual comparison is given in terms of graphs of the PDF. Lastly, snapshots of the time dependent MC and CDF solution in terms of the mean and two standard deviation uncertainty bounds are compared.

For the MC-data, the mean, standard deviation and skewness were calculated with built-in Matlab functions. For the CDF-solution, we used the following formulae:
\begin{align}
    \mu&=\bE[v] &&\text{(Mean)},\\
    \sigma&=\sqrt{\bE[(v-\mu)^2]}&&\text{(Standard deviation)},\\
    \text{Skew}&=\bE\left[\left(\frac{v-\mu}{\sigma}\right)^3\right]&&\text{(Skewness)},
\end{align}
where
\be
    \bE[g(v)]=\int_{V_{\text{min}}}^{V_{\text{max}}}g(x) f_v(x)dx.
\ee

Table \ref{tab:Results_CDFvsMC2} tabulates the statistical moments. Both the MC- and CDF-solution satisfy the property that the PDF integrates to unity.  
The other moments agree well, except for the skewness for a normal distribution of $a$, which should be zero. We attribute the deviation from zero to the regularization, which has a more severe effect on the left tail (see the remark below), thereby creating a slightly skewed approximation of a Gaussian. Overall, the agreement between MC and CDF-results is excellent, verifying consistency between MC and CDF.

\begin{table}
\centering
\caption{Statistical moments of the CDF solution with $N_x$=$N_V$=400 and the MC solution with $N_x$=100, $N_s=20,000$ and the optimal bandwidth, both at $t=5\cdot 10^{-3}$ and $x=0.03$.}
\begin{tabular}{|l|l|r|r|r|r|}
\hline 
\multicolumn{1}{|c|}{Distr.} & \multicolumn{1}{c|}{Method} & \multicolumn{1}{c|}{$\int f_v$} & \multicolumn{1}{c|}{$\mu$} & \multicolumn{1}{c|}{$\sigma$} & \multicolumn{1}{c|}{Skewness} \\ \hline
\multirow{2}{*}{Unif}        & MC                          & 1.000                              & 1.1867                    & 0.0532                   & -0.0061                       \\
                             & CDF                         & 1.000                             & 1.1872                    & 0.0530                   & -0.0080                       \\ \hline
\multirow{2}{*}{Norm}        & MC                          & 1.000                             & 1.1870                    & 0.0275                   & 0.0049                        \\
                             & CDF                         & 1.000                             & 1.1868                    & 0.0281                   & 0.0350                      \\ \hline
\multirow{2}{*}{Beta}        & MC                          & 1.000                             & 1.1471                    & 0.0296                   & 0.5904                        \\
                             & CDF                         & 1.000                             & 1.1477                    & 0.0301                    & 0.5681                       \\ \hline
\end{tabular}
\label{tab:Results_CDFvsMC2}
\end{table}

\begin{figure}
	\centering
    \subfigure[$a\sim\cU({[0.5,1.5]})$]{\label{fig:Results_CDFvsMC_Unif}\includegraphics[width=\textwidth]{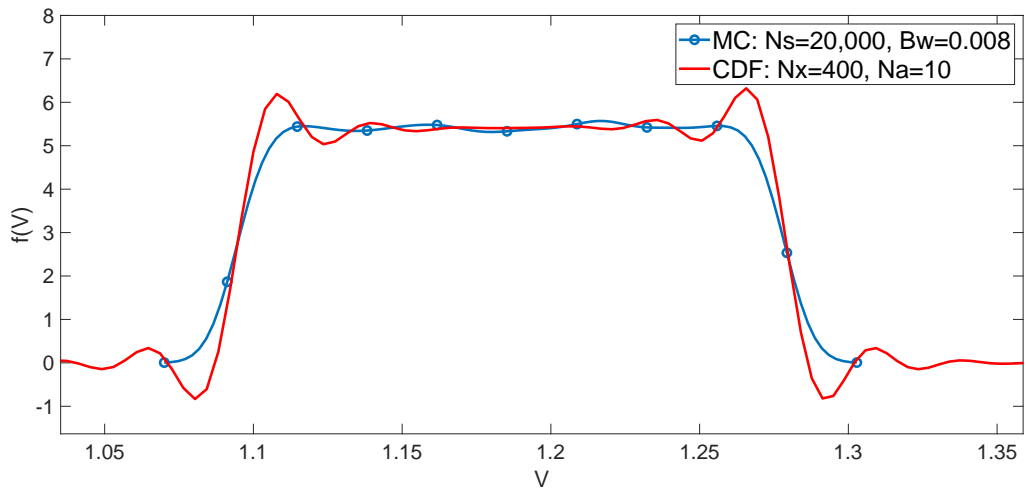}}
    \subfigure[$a\sim\cN(1,0.15)$]{\label{fig:Results_CDFvsMC_Norm}\includegraphics[width=\textwidth]{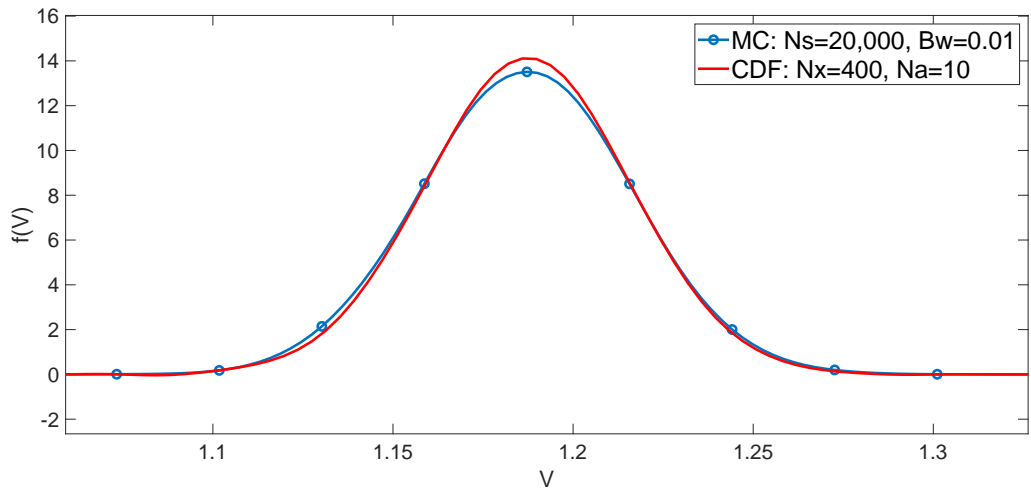}}
    \caption{PDF $f_v(V)$ at $t=5\cdot 10^{-3}$ and $x=0.03$ obtained by solving the CDF-equation (\ref{eq:setup_CDFequation}) and by MC simulation, assuming different distributions for $a$.}
\end{figure}

\begin{figure} 
	\centering
    \subfigure[$a\sim\cB(2,5)$]{\label{fig:Results_CDFvsMC_Beta}\includegraphics[width=\textwidth]{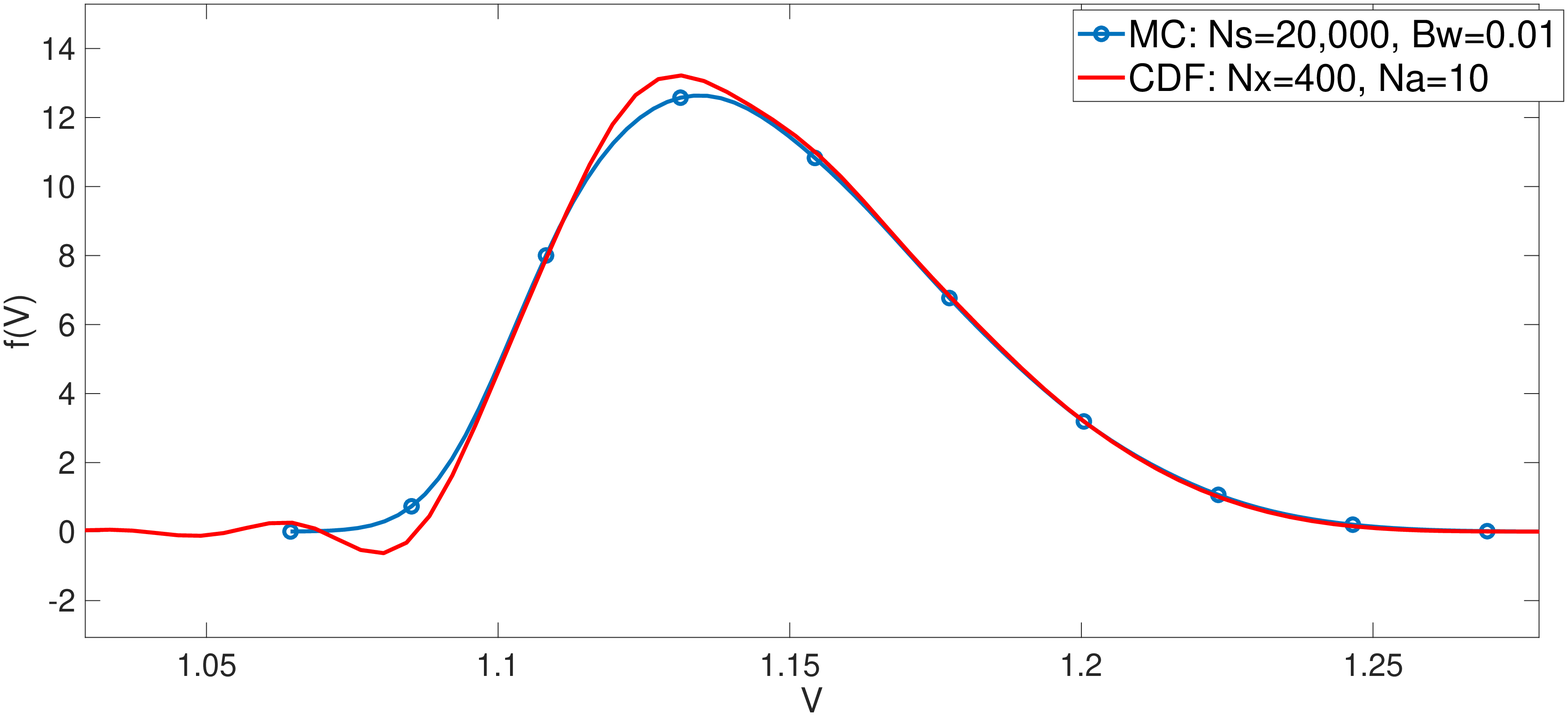}\addtocounter{subfigure}{2}}
    \caption{PDF $f_v(V)$ at $t=5\cdot 10^{-3}$ and $x=0.03$ obtained by solving the CDF-equation (\ref{eq:setup_CDFequation}) and by MC simulation, assuming different distributions for $a$.}
    \label{fig:Results_CDFvsMC}
\end{figure}

\begin{figure}
	\centering
    \includegraphics[width=\textwidth]{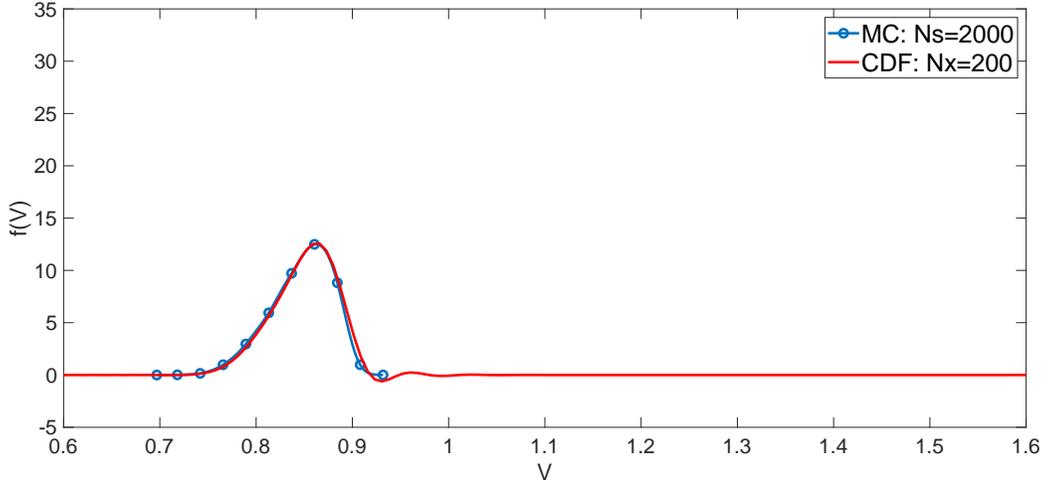}
    \caption{PDF $f_v(V)$ at $t=5\cdot 10^{-3}$ and $x=0.03$ obtained by solving the CDF-equation (\ref{eq:Remarks_CDFequation_NegativeSource}) and MC simulation, assuming a negative source and $a\sim\cB(2,5)$.}
    \label{fig:Remarks_CDF200vsMC2000_NegSource}
\end{figure}

Figure \ref{fig:Results_CDFvsMC} compares the PDFs determined with MC and CDF.
Overall the comparison is very good, particularly where the distribution is
smooth (Fig. \ref{fig:Results_CDFvsMC_Norm}-\ref{fig:Results_CDFvsMC_Beta}).
General features like support, position of the maximum (if applicable) and
general shape are in excellent agreement as can be expected from the  agreement
in the statistical moments. Initial undershoots and overshoots induced by
initial filtering with the non-positive delta kernel, have nearly disappeared
in the case of a normal distributed source coefficient. They are, however,
causing small deviations at the tails and tops of the normal and beta
distribution in Figure \ref{fig:Results_CDFvsMC_Norm} and
\ref{fig:Results_CDFvsMC_Beta}. In the case of a uniform (hence discontinuous)
distribution, the undershoots and overshoots are necessary for regularization
and apparent at the discontinuous edges of the distribution function. Away from
the regularization zones, however, the agreement is very good.
Figure \ref{fig:Remarks_CDF200vsMC2000_NegSource} shows the PDF for a negative steady Gaussian source
\be
    u(x)=-\tfrac{1}{\sqrt{2\pi}\sigma}e^{\frac{(x-x_a)^2}{2\sigma^2}}
\ee
and a beta-distributed $a$, is in excellent agreement between CDF and MC. Other distributions for $a$ give similar results, establishing consistency for negative sources. Only $N_x$=200 spatial intervals were considered for the CDF-equation, and $N_s$=2000 samples for the MC routine (all with the optimal bandwidth as mentioned before). Nevertheless, this is sufficient for the purpose of veryfication of the adjusted CDF-equation (\ref{eq:Remarks_CDFequation_NegativeSource}) for negative sources.

\begin{figure} 
	\centering
    \subfigure[$a\sim\cB(2,5)$]{\label{fig:Results_CDFvsMC_Beta_Mean_pm_2std_Allx}\includegraphics[width=\textwidth]{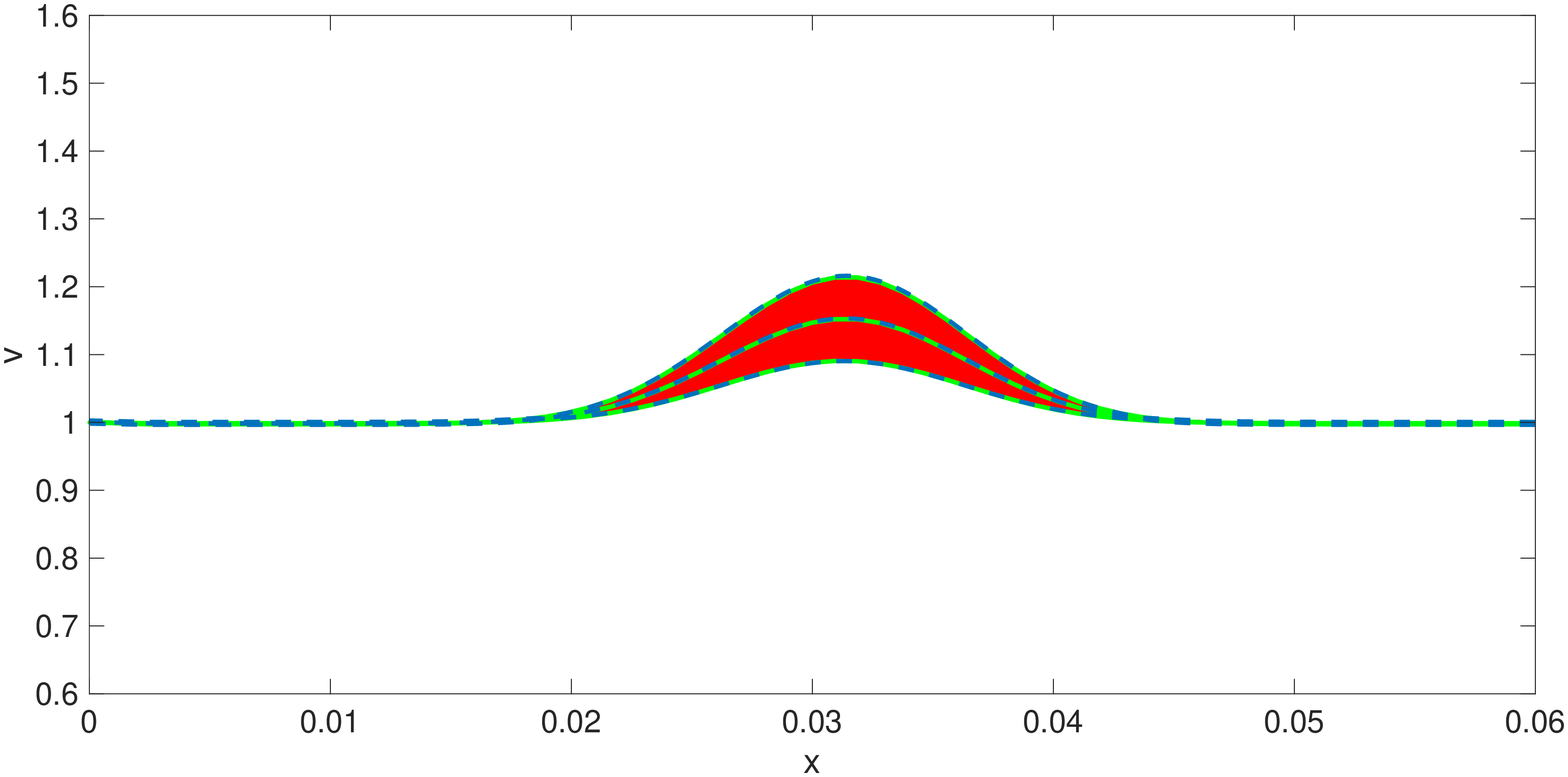}\addtocounter{subfigure}{2}}
    \caption{Two standard deviation uncertainty bounds for the velocity at $t=5\cdot 10^{-3}$, assuming different distributions for $a$. Green solid lines: MC. Blue dotted lines: CDF. Centerline is the mean. Colored version online.}
    \label{fig:Results_CDFvsMC_Mean_pm_2std_Allx}
\end{figure}

In Figure \ref{fig:Results_CDFvsMC_Mean_pm_2std_Allx}, two standard deviation uncertainty bounds for the velocity are plotted, to show the variability of the solution in the $x$-coordinate system. The uncertainty bounds are generated by calculating the standard deviation (in the way explained above) at every spatial grid point. 
Again, the agreement between CDF and MC in terms of mean and standard deviation is excellent and agrees in the eyeball norm.
\begin{figure} 
	\centering
    \subfigure[$a\sim\cU({[0.5,1.5]})$]{\label{fig:Results_CDFvsMC_Unif_Mean_pm_2std_Allx}\includegraphics[width=\textwidth]{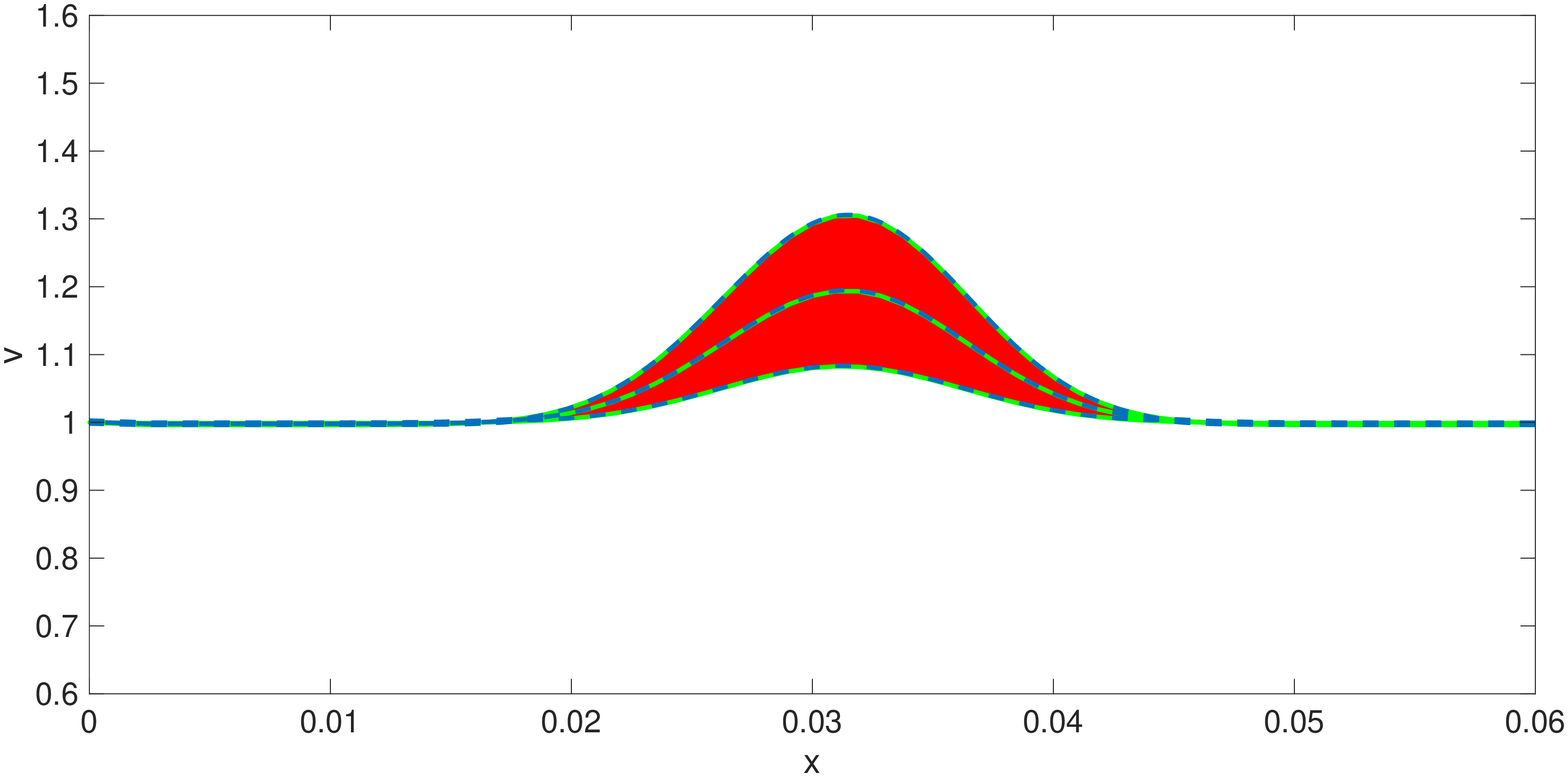}}
    \subfigure[$a\sim\cN(1,0.15)$]{\label{fig:Results_CDFvsMC_Norm_Mean_pm_2std_Allx}\includegraphics[width=\textwidth]{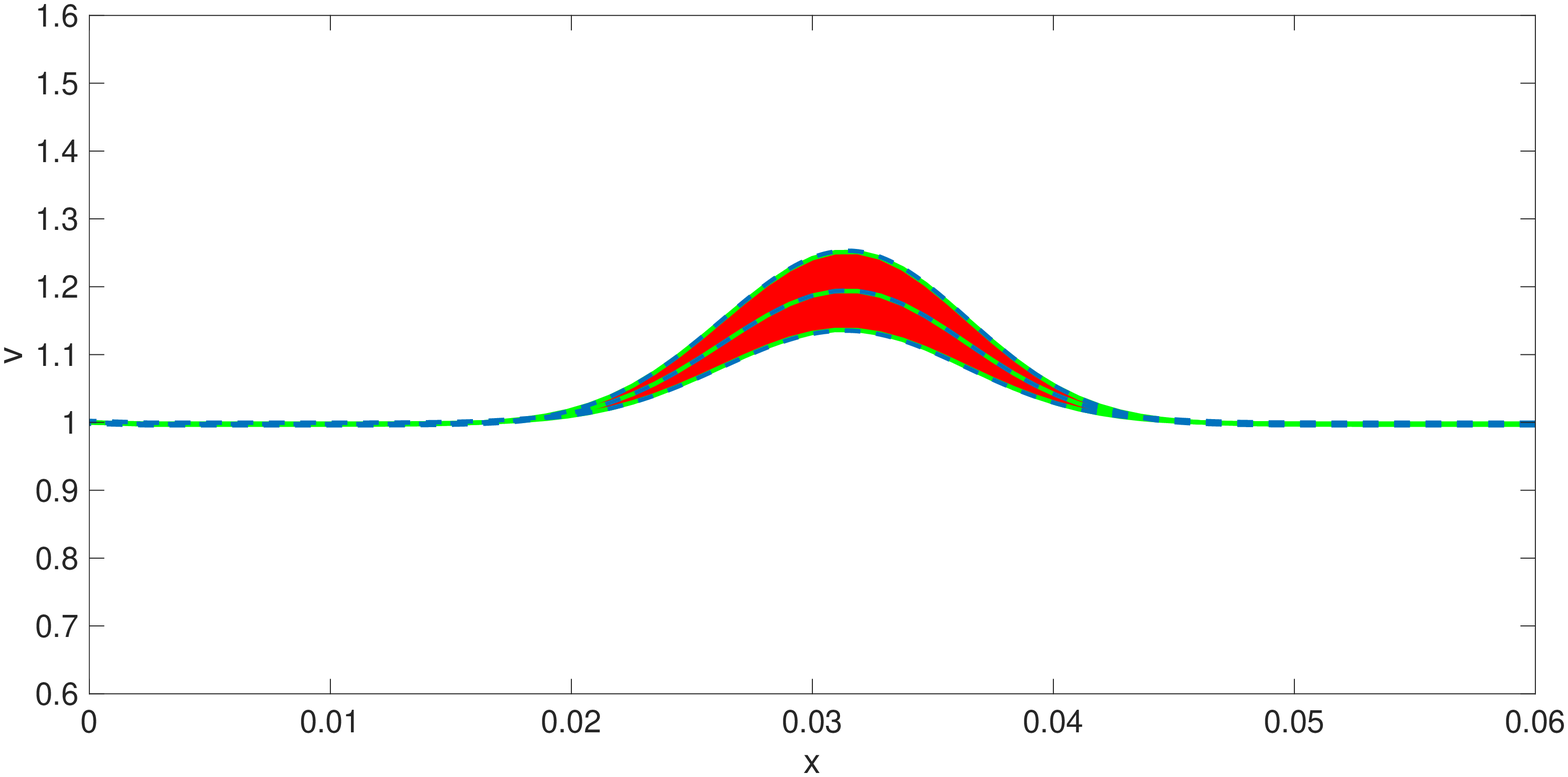}}
    \caption{Two standard deviation uncertainty bounds for the velocity at $t=5\cdot 10^{-3}$, assuming different distributions for $a$. Green solid lines: MC. Blue dotted lines: CDF. Centerline is the mean. Colored version online.}
\end{figure}   
\newline
\textit{Remark}: Because of the regularization of several and different singularities in CDF And MC, respectively, it is not possible to make rigorous statements about (the order of) convergence of the CDF-solution to the MC-solution as the number of ($x$- and $V$-)grid points increases; in the advected regularization zone, the CDF-solution does not converge to the MC-density, since there the former necessarily has undershoot and overshoot. For a uniformly distributed $a$ (Fig. \ref{fig:Results_CDFvsMC_Unif}), these zones lie at both boundaries of the support of the MC-solution. For normal and beta distributed $a$ (Fig. \ref{fig:Results_CDFvsMC_Norm} and \ref{fig:Results_CDFvsMC_Beta}), the propagated regularization zone runs from the left support boundary to the maximum of the MC-solution. It is clear from Figure \ref{fig:Results_CDFvsMC_Beta} that the right tail is less prone to under- and overshoot. Moreover, in the case of a uniform distributed coefficient the estimated kernel density does not have a flat plateau-shape (which the analytical solution should have), but shows multiple smaller peaks. We must therefore be satisfied with the eyeball norm comparisons, the good agreement that we generally find in the moments and the PDFs that confirm consistency between MC and CDF.




\section{Conclusions and Future Directions}
We have developed a CDF-model for systems with stochastic sources that may occur in models fornumerous multi-physics environments, including particle-laden flow and chemically reacting flow. In particular, Burgers' equation with random source coefficients has been considered, which renders the system stochastic. An equation for the joint CDF of the QoI and source coefficients is derived.

The major advantage of the CDF-approach over other existing methods - MC, SFEM, method of moments and PDF-equations - is that it provides a full description of the uncertain parameters, while being computationally efficient. Furthermore, the CDF-method results in a unambiguous, closed system of equations.

The initial condition in the CDF-equation contains a Heaviside function, which needs to be regularized to ensure stable solutions. We used a Dirac-delta polynomial kernel \cite{Suarez2014} for this. Numerical experiments show that a (purely numerical) boundary singularity appears in the CDF. An exponential filter \cite{Hesthaven} is applied to the solution after every time step to suppress this behaviour.

We compared solutions of a simplified CDF-equation with a positive source and only one random coefficient, with MC simulation. Chebyshev collocation is used for the spatial discretization of the PDE's, and a suitable higher-order Runge-Kutta method is employed for time-marching.

The CDF-method is shown to accurately predict the mean and standard deviation of the QoI, and is able to approximate the PDF of the QoI outside of the regularization zone, preserving the main characteristics of the PDF. However, under/overshoots generated by the regularization process make a thorough error analysis difficult.

For negative sources, an adjusted version of the method is needed to avoid severe instabilities within a few time steps. Instead of the joint CDF $F_{av}$, one can consider the variable $G_{av}=F_a-F_{av}$, which is not a CDF, but does satisfy the CDF-equation with adjusted initial and boundary conditions. Results are again in good agreement with MC simulation.

Future work will focus on extending the approach to more complex and general systems, like coupled gas-particle models, and to implementing a data-driven learning approach to decrease the uncertainty in the system.


\section*{Acknowledgements}
We thank Eindhoven University of Technology, and in particular Dr. Ten Thije Boonkkamp for arranging R.J.L. Rutjens' internship at SDSU. This research was supported by the Computational Mathematics program of AFOSR (grant FA9550-19-1-0387). 






\appendix
\section{Derivation of the Joint CDF Equation (Positive Source)}
\label{app:der}

In addition to the two random functions $\mathbf a$ and $v(x,t)$, we consider a fine-grained CDF
\begin{align}\label{eq:pi}
\Pi(\mathbf A, \mathbf a; V,v) \equiv \mathcal H(\mathbf A - \mathbf a) \mathcal H\left(V - v(x,t)\right),
\end{align}
where $\mathbf A$ and $V$ are deterministic variables, and $\mathcal H(\cdot)$ is the Heaviside function. The ensemble mean of any integrable function $g(\mathbf a,v)$ of random variables $\mathbf a \in \mathbb{R}^{N_g}$ and $v \in \mathbb{R}$ with the joint PDF $f_{\mathbf a v}(\mathbf A', V') : \mathbb{R}^{N_g} \times \mathbb{R} \rightarrow \mathbb{R}^+$ is
\begin{align}\label{mean-def}
\mathbb E[g(\mathbf a,v)] =  \int_\mathbb{R} \int_{\mathbb{R}^{N_g}} g(\mathbf A',V') f_{\mathbf a v}(\mathbf A',V') \text d \mathbf A' \text dV'.
\end{align}
In particular, at any space-time point $(x,t)$, the ensemble mean of $\Pi(\mathbf A, \mathbf a; V,v)$ over random realizations of the random variables $\mathbf a$ and $v$ is
\begin{align}\label{mean-pi}
\mathbb E[ \Pi ] = & \int_\mathbb{R} \int_{\mathbb{R}^{N_g}} \Pi(\mathbf A, \mathbf A'; V,V') f_{\mathbf a v}(\mathbf A',V';x,t) \text d \mathbf A' \text dV' \nonumber\\
=  & \int_\mathbb{R} \int_{\mathbb{R}^{N_g}} \mathcal H[\mathbf A - \mathbf A'] \mathcal H[V - V'] f_{\mathbf a v}(\mathbf A',V';x,t) \text d \mathbf A' \text dV'
\nonumber\\
= & \int_{-\infty}^V \int_{-\infty}^{A_1} \cdots \int_{-\infty}^{A_{N_g}}  f_{\mathbf a v}(\mathbf A',V';x,t) \text d \mathbf A' \text dV'= F_{\mathbf a v}(\mathbf A,V;x,t),
\end{align}
where $F_{\mathbf a v}$ is the joint cumulative distribution function (CDF) for $\mathbf a = \{a_0,\ldots,a_{N_g}\}$ and $v$ at any space-time point $(x,t)$. This property suggests a two-step procedure for the derivation of a PDE for $F_{\mathbf a v}$. First, derive a PDE for $\Pi$. Then, average this PDE. It follows from~\eqref{eq:pi} and the sifting property of the Dirac-delta function that
\begin{align}\label{eq:App_Der1_Pi_Properties}
\frac{\partial \Pi}{\partial t} = - \frac{\partial \Pi}{\partial V} \frac{\partial v}{\partial t}, \qquad
\frac{\partial \Pi}{\partial x} = - \frac{\partial \Pi}{\partial V} \frac{\partial v}{\partial x} \qquad\text{and}\qquad
g(v) \frac{\partial \Pi}{\partial V} = g(V) \frac{\partial \Pi}{\partial V},
\end{align}
for any test function $g(v)$. Hence, multiplication of the stochastic Burger's equation
\be
    \pd[v]{t}+v\pd[v]{x}=(u-v)\sum_{i=0}^{N_g}a_i\ChebT{i}(u-v)
\ee
with $- \partial \Pi / \partial V$ yields
\begin{equation}
 \frac{\partial \Pi}{\partial t} + V \frac{\partial \Pi}{\partial x} = - \frac{\partial \Pi}{\partial V}  (u-V) \sum_{i=0}^{N_g} a_i T_i(u-V).
\label{er}
\end{equation}
This is a linear PDE with the random (constant) coefficients $a_0.\cdots,a_{N_g}$. (Recall that both $u$ and $V$ are deterministic.) By virtue of~\eqref{mean-def} and~\eqref{mean-pi}, the ensemble average of this PDE is
\begin{align}
 \frac{\partial F_{\mathbf a v}}{\partial t} + V \frac{\partial F_{\mathbf a v} }{\partial x} = & - \int_\mathbb{R} \int_{\mathbb{R}^{N_g}} \frac{\partial \Pi}{\partial V}  (u-V) \sum_{i=0}^{N_g} A_i' T_i(u-V) f_{\mathbf a v}(\mathbf A',V';x,t) \text d \mathbf A' \text dV'
 \nonumber\\
 = & - (u-V) \sum_{i=0}^{N_g} \left[\int_\mathbb{R} \int_{\mathbb{R}^{N_g}} \frac{\partial \Pi}{\partial V} A_i'  f_{\mathbf a v}(\mathbf A',V';x,t) \text d \mathbf A' \text dV' \right] T_i(u-V)
  \nonumber\\
 = & - (u-V) \sum_{i=0}^{N_g} \left[ \frac{\partial }{\partial V} \int_\mathbb{R} \int_{\mathbb{R}^{N_g}} \Pi A_i'  f_{\mathbf a v}(\mathbf A',V';x,t) \text d \mathbf A' \text dV' \right] T_i(u-V). \label{eq:test1}
\end{align}
For the left hand side, we used the following theorem and lemma. We will omit the proofs.\\\\
\textbf{Theorem: (Dominated convergence)} \textit{Let $X_n$ be a sequence of integrable random variables and let the limit\\ $\lim_{n\to\infty}X_n(\omega)=X(\omega)$ exist for all $\omega\in\Omega$. If there is a nonnegative random variable $Y$ such that\\ $\vert X_n(\omega)\vert\leq Y(\omega)$ for all $\omega\in\Omega$ and all $n$, then $X$ is integrable and $\lim_{n\to\infty} \bE[X_n]=\bE[X]$.}\\\\
\textbf{Lemma:} \textit{Let $X\in\cX$ be a random variable and $g:\bR\times\cX\to\bR$ a function such that $g(X,t)$ is integrable for all $t$ and $g$ is differentiable with respect to $t$. Assume that there is a random variable $Y$ such that $\vert \pd{t}g(X,t)\vert\leq Y$ a.s. for all $t$, and $\bE[Y]<\infty$. Then $\pd{t}\bE[g(X,t)]=\bE[\pd{t}g(X,t)]$.}\\\\
Applied to our case, it thus follows that
\be
    \bE\left[\frac{\partial \Pi}{\partial t} + V \frac{\partial \Pi}{\partial x}\right] = \pd[\bE(\Pi)]{t}+V\pd[\bE(\Pi)]{x} = \frac{\partial F_{\mathbf a v}}{\partial t} + V \frac{\partial F_{\mathbf a v} }{\partial x}.
\ee
For the right hand side of Equation (\ref{eq:test1}), we interchange summation and integration and apply Leibniz' integral rule to obtain the final result.\\\\
The definition of $\Pi$ in terms of the Heaviside function implies that the integral in the last line on the right-hand-side of Equation (\ref{eq:test1}) reduces to
\begin{align}
\mathcal I = & \int_{-\infty}^V \int_{-\infty}^{A_1} \cdots \int_{-\infty}^{A_{N_g}}  A_i' f_{\mathbf a v}(\mathbf A',V';x,t) \text d \mathbf A' \text dV'.
\end{align}
Recalling the relationship between PDF and CDF, this can be written as
\begin{align}
\mathcal I  = & \int_{-\infty}^V \int_{-\infty}^{A_1} \cdots \int_{-\infty}^{A_{N_g}}  A_i' \frac{\partial^{N_g + 1} F_{\mathbf a v} }{ \partial A_1' \cdots \partial A_{N_g}' \partial V'}(\mathbf A',V';x,t) \text d \mathbf A' \text dV'
\nonumber\\
= & \int_{-\infty}^{A_i}  A_i' \frac{\partial F_{\mathbf a v} }{ \partial A_i' }(\mathbf A\!\setminus\! A_i, A_i',V;x,t) \text d  A_i'
\nonumber\\
= & A_i F_{\mathbf a v} (\mathbf A,V;x,t) - \int\limits_{-\infty}^{A_i} F_{\mathbf a v} (\mathbf A\!\setminus\! A_i, A_i',V;x,t) \text d A_i',
\end{align}
where
\be
    \vA\!\setminus\! A_i = \left(A_1,\ldots,A_{i-1},A_{i+1},\ldots,A_{N_g}\right).
\ee
Putting things back together leads to
\be
    \frac{\partial
	F_{\mathbf a v}}
	{\partial t} + V \frac{\partial F_{\mathbf a v}}{\partial x} = - (u-V) \sum_{i=0}^{N_g} T_i(u-V) \frac{\partial }{\partial V} \left[  A_i F_{\mathbf a v} - \int\limits_{-\infty}^{A_i} F_{\mathbf a v} (\mathbf A\!\setminus\! A_i, A_i',V;x,t) \text d A_i' \right] .
\ee

\bibliographystyle{IJ4UQ_Bibliography_Style}

\bibliography{References_Rik}

\begin{thebibliography}{10}

\bibitem{Najm2009}
Najm, H., Debusschere, B., Marzouk, Y., Widmer, S., and {Le Ma{\^{i}}tre}, O.,
  {Uncertainty quantification in chemical systems}, {\em Int. J. Numer. Meth.
  Engng}, 80(6-7):789--814, 2009.

\bibitem{Cai2016}
Cai, L., Pitsch, H., Mohamed, S.Y., Raman, V., Bugler, J., Curran, H., and
  Sarathy, S.M., {Optimized reaction mechanism rate rules for ignition of
  normal alkanes}, {\em Combustion and Flame}, 173:468--482, 2016.

\bibitem{Reagan2005}
Reagan, M.T., Najm, H.N., Pébay, P.P., Knio, O.M., and Ghanem, R.G.,
  Quantifying uncertainty in chemical systems modeling, {\em International
  Journal of Chemical Kinetics}, 37(6):368--382, 2005.

\bibitem{boso2018}
Boso, F., Marzadri, A., and Tartakovsky, D.M., Probabilistic forecasting of
  nitrogen dynamics in hyporheic zone, {\em Water Resour. Res.},
  54(7):4417--4431, 2018.

\bibitem{Fountoulakis2018}
Fountoulakis, V., Udaykumar, H.S., and Jacobs, G.B., {Uncertainty
  quantification in Eulerian-Lagrangian simulations of (point-)particle-laden
  flows with data-driven and empirical forcing models}, {\em arXiv e-prints},
  2018, arXiv:1810.13047.

\bibitem{Shotorban2013}
Shotorban, B., Jacobs, G.B., Ortiz, O., and Truong, Q., {An Eulerian model for
  particles nonisothermally carried by a compressible fluid}, {\em
  International Journal of Heat and Mass Transfer}, 65:845--854, 2013.

\bibitem{Guerra2016}
Guerra, G.M., Zio, S., Camata, J.J., Dias, J., Elias, R.N., Mattoso, M., Paulo,
  P.L., Alvaro, A.L., and Rochinha, F.A., {Uncertainty quantification in
  numerical simulation of particle-laden flows}, {\em Computational
  Geosciences}, 20(1):265--281, 2016.

\bibitem{Giles2008}
Giles, M.B., {Multilevel Monte Carlo path simulation}, {\em Operations
  Research}, 56(3):607--617, 2008.

\bibitem{Giles2013}
{Giles}, M.B., {Multilevel Monte Carlo methods}, {\em arXiv e-prints}, April
  2013, arXiv:1304.5472.

\bibitem{Iman1980}
Iman, R.L. and Conover, W., Small sample sensitivity analysis techniques for
  computer models with an application to risk assessment, {\em Communications
  in Statistics - Theory and Methods}, 9(17):1749--1842, 1980.

\bibitem{Owen2013}
Owen, A.B.
\newblock {\em Monte Carlo theory, methods and examples}, chapter~10.
\newblock 2013.
\newblock Accessed 15 May 2019, from
  \url{https://statweb.stanford.edu/~owen/mc/Ch-var-adv.pdf}.

\bibitem{Wiener1938}
Wiener, N., The homogeneous chaos, {\em American Journal of Mathematics},
  60(4):897--936, 1938.

\bibitem{Stefanou2009}
Stefanou, G., {The stochastic finite element method : past, present and
  future}, {\em Comput. Methods Appl. Mech. Engrg.}, 198(9-12):1031--1051,
  2009.

\bibitem{Wan2005}
Wan, X. and Karniadakis, G.E., An adaptive multi-element generalized polynomial
  chaos method for stochastic differential equations, {\em Journal of
  Computational Physics}, 209(2):617--642, 2005.

\bibitem{Wan2006}
Wan, X. and Karniadakis, G.E., Multi-element generalized polynomial chaos for
  arbitrary probability measures, {\em SIAM J. Sci. Comput.}, 28(3):901--928,
  March 2006.

\bibitem{Debusschere2017}
Debusschere, B.
\newblock {Intrusive Polynomial Chaos Methods for Forward Uncertainty
  Propagation}.
\newblock In {\em Handbook of Uncertainty Quantification\/} \cite{Ghanem2017},
  chapter~17, pp. 617--636.

\bibitem{Xiu2016}
Xiu, D.
\newblock Stochastic collocation methods: a survey.
\newblock In {\em Handbook of Uncertainty Quantification\/} \cite{Ghanem2017},
  chapter~20, pp. 699--716.

\bibitem{Ma2009}
Ma, X. and Zabaras, N., An adaptive hierarchical sparse grid collocation
  algorithm for the solution of stochastic differential equations, {\em Journal
  of Computational Physics}, 228(8):3084--3113, 2009.

\bibitem{Barajas2016}
Barajas-Solano, D. and Tartakovsky, D., Stochastic collocation methods for
  nonlinear parabolic equations with random coefficients, {\em SIAM/ASA Journal
  on Uncertainty Quantification}, 4(1):475--494, 2016.

\bibitem{Pope2000}
Pope, S., {\em {Turbulent flows}}, Cambridge University Press, Cambridge, 2000.

\bibitem{lichtner-2003-stochastic}
Lichtner, P.C. and Tartakovsky, D.M., Stochastic analysis of effective rate
  constant for heterogeneous reactions, {\em Stoch. Environ. Res. Risk
  Assess.}, 17(6):419--429, 2003.

\bibitem{tartakovsky-2009-probability}
Tartakovsky, D.M., Dentz, M., and Lichtner, P.C., Probability density functions
  for advective-reactive transport in porous media with uncertain reaction
  rates, {\em Water Resour. Res.}, 45:W07414, 2009.

\bibitem{broyda-2010-probability}
Broyda, S., Dentz, M., and Tartakovsky, D.M., Probability density functions for
  advective-reactive transport in radial flow, {\em Stoch. Environ. Res. Risk
  Assess.}, 24(7):985--992, 2010.

\bibitem{dentz-2010-probability}
Dentz, M. and Tartakovsky, D.M., Probability density functions for passive
  scalars dispersed in random velocity fields, {\em Geophys. Res. Lett.},
  37:L24406, 2010.

\bibitem{wang-2013-probability}
Wang, P., Tartakovsky, A.M., and Tartakovsky, D.M., Probability density
  function method for {Langevin} equations with colored noise, {\em Phys. Rev.
  Lett.}, 110(14):140602, 2013.

\bibitem{Suarez2014}
Suarez, J., Jacobs, G.B., and Don, W.S., {A high-order Dirac-delta
  regularization with optimal scaling in the spectral solution of
  one-dimensional singular hyperbolic conservation laws}, {\em Journal of
  Scientific Computing}, 36(4):A1831--A1849, 2014.

\bibitem{Wissink2018}
Wissink, B.W., Jacobs, G.B., Ryan, J.K., Don, W.S., and van~der Weide, E.T.A.,
  {Shock regularization with smoothness-increasing accuracy-conserving
  Dirac-delta polynomial kernels}, {\em Journal of Scientific Computing},
  77(1):579--596, 2018.

\bibitem{Gottlieb2001}
Gottlieb, D. and Hesthaven, J.S., {Spectral methods for hyperbolic problems},
  {\em Journal of Computational and Applied Mathematics}, 128:83--131, 2001.

\bibitem{Hesthaven}
Hesthaven, J.S., Gottlieb, S., and Gottlieb, D., {\em Spectral methods for
  time-dependent problems}, Cambridge Monographs on Applied and Computational
  Mathematics, Cambridge University Press, Cambridge, 2007.

\bibitem{Press1992}
Press, W.H., Teukolsky, S.A., Vetterling, W.T., and Flannery, B.P.
\newblock {\em {Numerical Recipes for Fortran 77: The Art of Scientific
  Computing}}, p. 715.
\newblock Cambridge University Press, 2nd edition, 1992.

\bibitem{Gottlieb1998}
Gottlieb, S. and Shu, C., {Total variation diminishing Runge-Kutta schemes},
  {\em Math.Comput.}, 67(221):73--85, 1998.

\bibitem{Vandeven1991}
Vandeven, H., Family of spectral filters for discontinuous problems, {\em
  Journal of Scientific Computing}, 6(2):159--192, Jun 1991.

\bibitem{ksdensity}
{The MathWorks, Inc.}
\newblock {\em {Matlab Statistics and Machine Learning Toolbox User's Guide
  R2019a}}, pp. 3603--3629.
\newblock 2019.

\bibitem{Silverman1986}
Silverman, B.W., {\em {Density estimation for statistics and data analysis}},
  Monographs on Statistics and Applied Probability, Chapman and Hall, London,
  1986.

\bibitem{Ghanem2017}
Ghanem, R., Owhadi, H., and Higdon, D., {\em Handbook of Uncertainty
  Quantification}, Springer Switzerland, 2017.

\end{thebibliography}
\end{document}